\documentclass[twocolumn,american,aps,prb]{revtex4}
\usepackage{mathptmx}

\usepackage[T1]{fontenc}
\usepackage[latin9]{inputenc}
\usepackage{textcomp}
\usepackage{amsmath}
\usepackage{amssymb}
\usepackage{graphicx}

\makeatletter
\@ifundefined{textcolor}{}
{%
 \definecolor{BLACK}{gray}{0}
 \definecolor{WHITE}{gray}{1}
 \definecolor{RED}{rgb}{1,0,0}
 \definecolor{GREEN}{rgb}{0,1,0}
 \definecolor{BLUE}{rgb}{0,0,1}
 \definecolor{CYAN}{cmyk}{1,0,0,0}
 \definecolor{MAGENTA}{cmyk}{0,1,0,0}
 \definecolor{YELLOW}{cmyk}{0,0,1,0}
}

\usepackage{float}

\usepackage{textcomp}

\usepackage{esint}



\makeatother

\usepackage{babel}
\begin{document}

\title{\noindent {\Large Physics of self-sustained oscillations in the positive
glow corona }}

\author{\noindent \textbf{Sung Nae Cho}}

\email{sungnae.cho@samsung.com }

\affiliation{\noindent Micro Devices Group, Micro Systems Laboratory, Samsung
Advanced Institute of Technology, Samsung Electronics Co., Ltd, Mt.
14-1 Nongseo-dong, Giheung-gu, Yongin-si, Gyeonggi-do 446-712, Republic
of Korea. }

\date{5 July 2012 }
\begin{abstract}
\noindent The physics of self-sustained oscillations in the phenomenon
of positive glow corona is presented. The dynamics of charged-particle
oscillation under static electric field has been briefly outlined;
and, the resulting self-sustained current oscillations in the electrodes
have been compared with the measurements from the positive glow corona
experiments. The profile of self-sustained electrode current oscillations
predicted by the presented theory qualitatively agrees with the experimental
measurements. For instance, the experimentally observed saw-tooth
shaped electrode current pulses are reproduced by the presented theory.
Further, the theory correctly predicts the pulses of radiation accompanying
the abrupt rises in the saw-tooth shaped current oscillations, as
verified from the various glow corona experiments.\textbf{ }
\end{abstract}
\maketitle

\section{Introduction}

\begin{figure}[h]
\begin{centering}
\includegraphics[width=0.9\columnwidth]{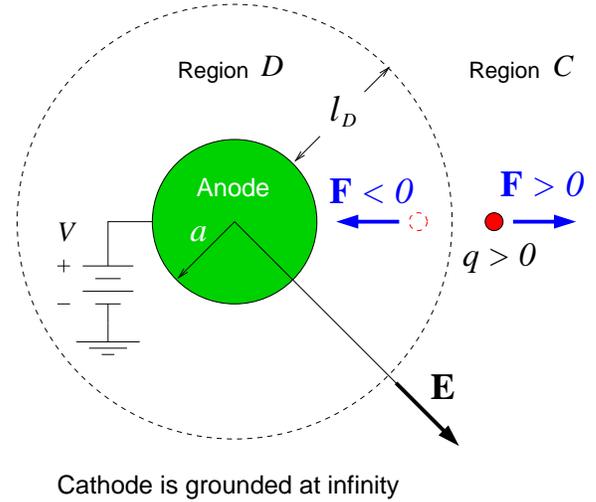}
\par\end{centering}

\caption{(Color online) Charged point particle near a conducting sphere of
radius $a,$ fixed at voltage $V>0.$ In region $D,$ the particle
is attracted to the sphere whereas, in region $C,$ the particle is
repulsed from the sphere. The electric field, $\mathbf{E},$ is in
the radially outward direction. \label{fig:1}}
\end{figure}

Consider a point particle with charge $q>0$ near the conducting sphere
of radius $a$ and fixed at potential $V>0,$ which is illustrated
in Fig. \ref{fig:1}. The force acting on the charged point particle
is given by\cite{J. D. Jackson} 
\begin{equation}
\mathbf{F}=qa\left[\frac{V}{r^{3}}-\frac{q}{4\pi\epsilon_{o}\left(r^{2}-a^{2}\right)^{2}}\right]\mathbf{R},\label{eq:F-Jackson}
\end{equation}
 where $\mathbf{R}$ is the particle's position vector, $\epsilon_{o}$
is the vacuum permittivity, and $r$ is the radial length from sphere's
center. In Eq.  (\ref{eq:F-Jackson}), the force becomes negative
in the limit $r$ approaches $a$ and becomes positive in the limit
$r$ goes to infinity. Between these two limits, there is a point
of unstable equilibrium at which the force vanishes. Such location
is identified by $r=a+l_{D}$ in Fig. \ref{fig:1}, which point represents
the borderline between regions $C$ and $D.$ In region $C,$ the
particle is repulsed from the sphere whereas, in region $D,$ the
particle is attracted to the sphere. Why? Well, nothing surprising
here. The positive point particle induces negative charges at the
sphere's surface; and, the force between the two is always attractive.
This attractive force dominates in region $D,$ and the point particle
is attracted to sphere there. 

\begin{figure}[h]
\begin{centering}
\includegraphics[width=0.8\columnwidth]{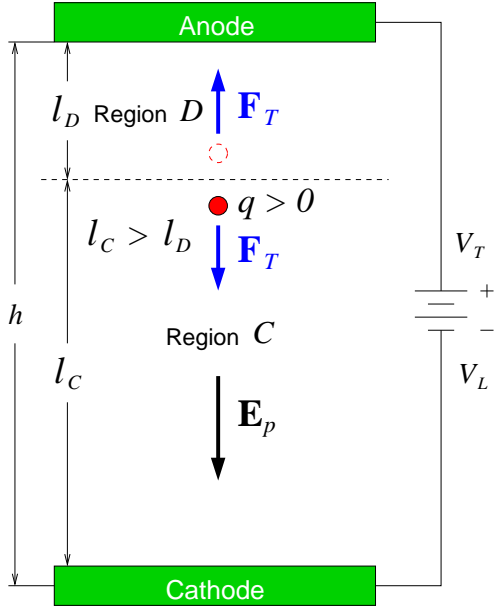}
\par\end{centering}

\caption{(Color online) Charged point particle inside a plane-parallel conductors
separated by $h.$ In region $D,$ the particle is attracted to the
anode whereas, in region $C,$ the particle is attracted to the cathode.
\label{fig:2}}
\end{figure}

The same physics can be applied to describe the behavior of a charged
point particle between the plane-parallel plates, which is illustrated
in Fig. \ref{fig:2}. At distances close to the anode, the charged
point particle is attracted to the anode's surface whereas, for all
other distances between the plates, the particle is repulsed in the
direction of the parallel plate electric field, $\mathbf{E}_{p},$
which field is present even in the absence of the charged-particle.
Consequently, the space between the plates is divided into regions
$C$ and $D,$ where the location of unstable equilibrium is at distance
$l_{D}$ from the surface of the anode, as indicated in Fig. \ref{fig:2}. 

\begin{figure*}[t]
\begin{centering}
\includegraphics[width=1.4\columnwidth]{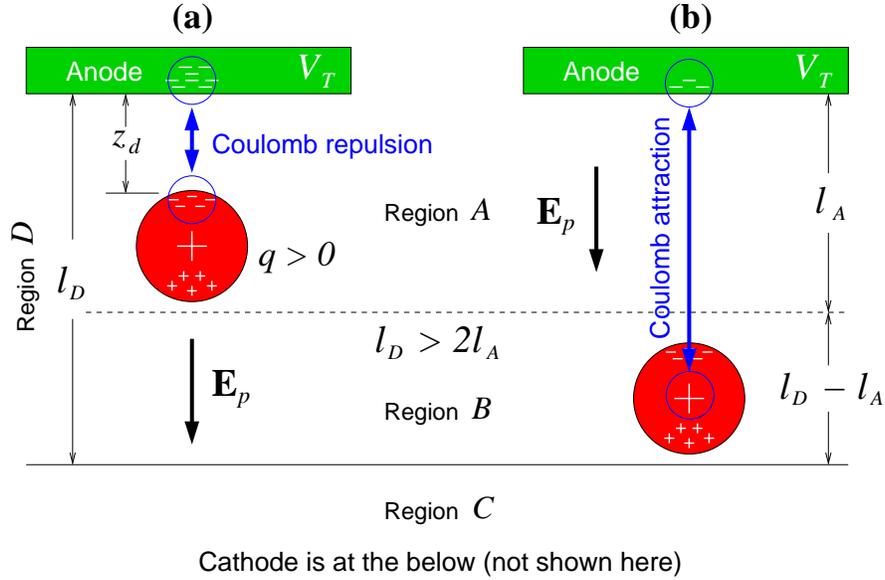}
\par\end{centering}

\caption{(Color online) (a) In case of a structured charged-particle, the region
$D$ in Fig. \ref{fig:2} is further divided into two regions $A$
and $B.$ The particle is repulsed from the anode in region $A$ due
to the Coulomb repulsion between the image charge at the anode's surface
and the negative bound charges formed at the surface of particle's
upper hemisphere. The surface of charged spherical particle is an
equipotential surface; and, any excess surface charges there are uniformly
distributed over it. Without confusion, the contributions from such
excess charges, uniformly distributed over the surface of particle,
are indicated by a big ``+'' symbol at the particle's center to
distinguish them from the positive bound charges associated with the
induced depolarization field, which are indicated by small ``+''
at the particle's lower surface. (b) The polarized charged-particle
is attracted to the anode in region $B$ due to the Coulomb attraction
arising between the particle's excess charge $q$ and the image (or
induced) charge associated with it at the anode's surface. The width
of region $A$ is identified by $l_{A}$ and the width of region $B$
is given by $l_{D}-l_{A},$ where $l_{D}$ is the borderline between
regions $D$ and $C.$ \label{fig:3}}
\end{figure*}

In both cases, the charged-particle dynamics is pretty boring. The
charged point particle ends up adhering to the surface of the anode
when it is in region $D$ whereas, when it is in region $C,$ the
particle gets repulsed in the direction of the applied electric field.
The problem becomes interesting when a charged-particle with structure
is considered. Unlike the point particle, the structured particle
can be polarized under externally applied electric field, such as
$\mathbf{E}$ and $\mathbf{E}_{p}$ in Figs. \ref{fig:1} and \ref{fig:2},
respectively. The resulting depolarization field formed inside of
the structured particle redistributes the negative bound charges to
the particle's upper hemisphere surface and leaves the particle's
lower hemisphere surface depleted of the negative bound charges. Such
redistribution of the bound charges inside of the structured particle
is schematically illustrated in Figs. \ref{fig:3}(a) and \ref{fig:3}(b).
As a consequence of the depolarization field inside of the structured
particle, the particle is repulsed from the surface of anode in region
$A$ due to the Coulomb repulsion arising between the negative charges
induced at the anode's surface and the negative bound charges formed
at the surface of the particle's upper hemisphere, as illustrated
in Fig. \ref{fig:3}(a). Such repulsive force decays rapidly outside
of the region $A.$ In region $B,$ the force acting on the particle
is dominated by the Coulomb attraction between the particle's excess
positive charge, $q>0,$ and the negative charges induced by it at
the surface of the anode. This force is eventually overwhelmed by
the Coulomb repulsion in region $A,$ and the whole process gets repeated,
thereby resulting in a charged-particle oscillation in region $D.$ 

The discussed charged-particle oscillation mechanism is intrinsic
to the phenomenon of glow corona.\cite{corona-discharge-1,corona-discharge-2,corona-discharge-3,corona-discharge-4,corona-discharge-5}
The self-sustained pulsing in the positive glow corona (also referred
to as DC glow discharge) can be qualitatively explained from the aforementioned,
oscillating, charged-particle dynamics. In this paper, I shall show
that the self-sustained pulsing in the positive glow corona involves
the kind of charged-particle oscillation mechanism discussed in Figs.
\ref{fig:3}(a) and \ref{fig:3}(b). To accomplish this, I shall,
first, briefly outline the dynamics of charged-particle oscillation
under constant electric field. The result is then used to predict
the current oscillations in the electrodes. This prediction of electrode
current oscillations is compared with the results from the various
glow corona experiments.

\section{\noindent Charged-particle oscillation in constant electric field}

\begin{figure}[h]
\begin{centering}
\includegraphics[width=1\columnwidth]{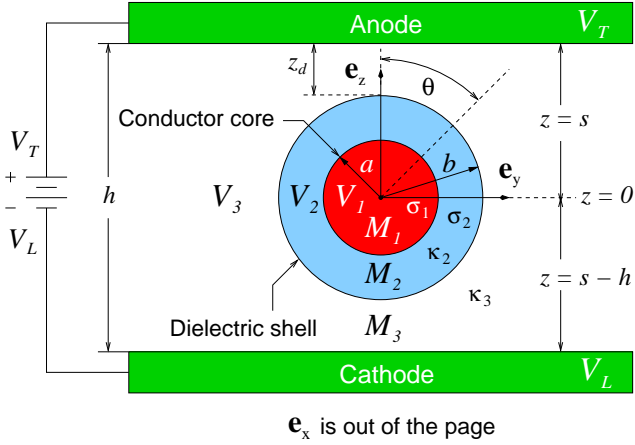}
\par\end{centering}

\caption{ (Color online) Configuration showing a core-shell structured charged-particle
between a DC voltage biased plane-parallel conductors. \label{fig:4}}
\end{figure}

The dynamics of charged-particle oscillation is discussed by considering
a model configuration illustrated in Fig. \ref{fig:4}, where a core-shell
structured particle has a conductive core of radius $a$ and an insulating
shell of thickness $b-a.$ The conductor core has a surface free charge
density of $\sigma_{1}$ and the insulator has a surface charge density
of $\sigma_{2}.$ The surface charge density on the insulator has
been introduced purely for mathematical generalization. In the final
expression, this term can be eliminated by setting $\sigma_{2}=0\,\textnormal{C}\cdot\textnormal{m}^{-2}.$
The $\kappa_{2}$ and $\kappa_{3}$ represent the dielectric constants
of the insulating shell and the space between the plates, respectively. 

The potential in regions $M_{1},$ $M_{2},$ and $M_{3}$ of Fig.
\ref{fig:4} are obtained by solving the Laplace equation, $\nabla^{2}V=0,$
with appropriate boundary conditions. Solutions $V_{1},$ $V_{2},$
and $V_{3}$ are given by\cite{Cho} 
\begin{align}
V_{1} & =V_{L}+\alpha+E_{p}\left(h-s\right),\quad r\leq a,\label{eq:V1-FINAL}
\end{align}

\begin{align}
V_{2}\left(r,\theta\right) & =V_{L}+\beta+E_{p}\left(h-s+\gamma r\cos\theta\right)\nonumber \\
 & -\frac{\lambda}{r}-\frac{a^{3}\gamma E_{p}\cos\theta}{r^{2}},\quad a<r\leq b,\label{eq:V2-FINAL}
\end{align}
 
\begin{align}
 & V_{3}\left(r,\theta\right)=V_{L}+E_{p}\left(h-s+r\cos\theta\right)+\frac{\nu}{r}\nonumber \\
 & \quad+\frac{\left[\gamma\left(b^{3}-a^{3}\right)-b^{3}\right]E_{p}\cos\theta}{r^{2}}+C,\quad r>b,\label{eq:V3-FINAL}
\end{align}
 where $C$ is a constant, $\theta$ is spherical polar angle defined
in Fig. \ref{fig:4},  $r$ is radial length, $E_{p}$ is parallel-plate
electric field, 
\begin{equation}
E_{p}\equiv\left\Vert \mathbf{E}_{p}\right\Vert =\frac{V_{T}-V_{L}}{h},\quad V_{T}>V_{L};\label{eq:Ep-DEF}
\end{equation}
 and, constants $\alpha,$ $\beta,$ $\gamma,$ $\lambda,$ and $\nu$
are defined as 

\begin{align}
\alpha & =\frac{a\left(b-a\right)\sigma_{1}}{b\epsilon_{0}\kappa_{2}}+\frac{a^{2}\sigma_{1}+b^{2}\sigma_{2}}{b\epsilon_{0}\kappa_{3}},\nonumber \\
\beta & =\frac{a\left(2b-a\right)\sigma_{1}}{b\epsilon_{0}\kappa_{2}}+\frac{a^{2}\sigma_{1}+b^{2}\sigma_{2}}{b\epsilon_{0}\kappa_{3}},\nonumber \\
\gamma & =\frac{3\kappa_{3}b^{3}}{\left(\kappa_{2}+2\kappa_{3}\right)b^{3}+2\left(\kappa_{2}-\kappa_{3}\right)a^{3}},\label{eq:alpla-beta-gama-lambda-nu}\\
\lambda & =\frac{a^{2}\sigma_{1}}{\epsilon_{0}\kappa_{2}},\nonumber \\
\nu & =\frac{2a\left(b-a\right)\sigma_{1}}{\epsilon_{0}\kappa_{2}}+\frac{a^{2}\sigma_{1}+b^{2}\sigma_{2}}{\epsilon_{0}\kappa_{3}}.\nonumber 
\end{align}

The electric displacement in region $M_{3}$ is obtained by computing
\[
\mathbf{D}_{3}\left(r,\theta\right)=-\epsilon_{0}\kappa_{3}\nabla V_{3}\left(r,\theta\right),
\]
 where 
\[
\nabla=\mathbf{e}_{r}\frac{\partial}{\partial r}+\mathbf{e}_{\theta}\frac{1}{r}\frac{\partial}{\partial\theta}+\mathbf{e}_{\phi}\frac{1}{r\sin\theta}\frac{\partial}{\partial\phi},
\]
 
\begin{align*}
\mathbf{e}_{r} & =\mathbf{e}_{x}\sin\theta\cos\phi+\mathbf{e}_{y}\sin\theta\sin\phi+\mathbf{e}_{z}\cos\theta,\\
\mathbf{e}_{\theta} & =\mathbf{e}_{x}\cos\theta\cos\phi+\mathbf{e}_{y}\cos\theta\sin\phi-\mathbf{e}_{z}\sin\theta,\\
\mathbf{e}_{\phi} & =-\mathbf{e}_{x}\sin\phi+\mathbf{e}_{y}\cos\phi.
\end{align*}
 It can be shown that $\mathbf{e}_{z}$ component of $\mathbf{D}\left(r,\theta\right)$
in region $M_{3},$ evaluated at the anode's surface, is\cite{Cho}
\begin{align*}
 & \mathbf{D}_{3;z}\left(x,y,s\right)=\mathbf{e}_{z}\epsilon_{0}\kappa_{3}\left\{ \frac{3\left[\gamma\left(b^{3}-a^{3}\right)-b^{3}\right]E_{p}s^{2}}{\left(x^{2}+y^{2}+s^{2}\right)^{5/2}}\right.\\
 & \left.+\frac{\nu s-\left[\gamma\left(b^{3}-a^{3}\right)-b^{3}\right]E_{p}}{\left(x^{2}+y^{2}+s^{2}\right)^{3/2}}-E_{p}\right\} .
\end{align*}
 At the anode's surface, the $\mathbf{e}_{z}$ component of $\mathbf{D}\left(x,y,z\right)$
suffers a discontinuity, 
\[
\mathbf{e}_{z}\cdot\mathbf{D}_{3;z}\left(x,y,s\right)=-\sigma_{iup};
\]
 and, the surface charge density, there, is 

\begin{align*}
\sigma_{iup}\left(\rho,s\right) & =-\epsilon_{0}\kappa_{3}\left\{ \frac{3\left[\gamma\left(b^{3}-a^{3}\right)-b^{3}\right]E_{p}s^{2}}{\left(\rho^{2}+s^{2}\right)^{5/2}}\right.\\
 & \left.+\frac{\nu s-\left[\gamma\left(b^{3}-a^{3}\right)-b^{3}\right]E_{p}}{\left(\rho^{2}+s^{2}\right)^{3/2}}-E_{p}\right\} ,
\end{align*}
where $\rho\equiv\sqrt{x^{2}+y^{2}}.$ Similarly, the $\mathbf{e}_{z}$
component of $\mathbf{D}\left(x,y,z\right)$ in region $M_{3},$ evaluated
at the cathode's surface, is 
\begin{align*}
 & \mathbf{D}_{3;z}\left(x,y,s-h\right)\\
 & \quad=\mathbf{e}_{z}\epsilon_{0}\kappa_{3}\left\{ \frac{3\left[\gamma\left(b^{3}-a^{3}\right)-b^{3}\right]E_{p}\left(s-h\right)^{2}}{\left[x^{2}+y^{2}+\left(s-h\right)^{2}\right]^{5/2}}\right.\\
 & \quad\left.+\frac{\nu\left(s-h\right)-\left[\gamma\left(b^{3}-a^{3}\right)-b^{3}\right]E_{p}}{\left[x^{2}+y^{2}+\left(s-h\right)^{2}\right]^{3/2}}-E_{p}\right\} .
\end{align*}
 At the cathode's surface, the $\mathbf{e}_{z}$ component of $\mathbf{D}\left(x,y,z\right)$
suffers a discontinuity, 
\[
\mathbf{e}_{z}\cdot\mathbf{D}_{3;z}\left(x,y,s-h\right)=\sigma_{ilp};
\]
 and, there, the surface charge density is 

\begin{align*}
\sigma_{ilp}\left(\rho,s\right) & =\epsilon_{0}\kappa_{3}\left\{ \frac{3\left[\gamma\left(b^{3}-a^{3}\right)-b^{3}\right]E_{p}\left(h-s\right)^{2}}{\left[\rho^{2}+\left(h-s\right)^{2}\right]^{5/2}}\right.\\
 & \left.-\frac{\nu\left(h-s\right)+\left[\gamma\left(b^{3}-a^{3}\right)-b^{3}\right]E_{p}}{\left[\rho^{2}+\left(h-s\right)^{2}\right]^{3/2}}-E_{p}\right\} .
\end{align*}

The force acting on the particle is\cite{Cho} 

\begin{align}
\mathbf{F}_{T} & =\mathbf{F}_{1}+\mathbf{F}_{2}-\mathbf{e}_{z}mg,\label{eq:FT-pre}
\end{align}
 where $m$ is particle's mass, $g=9.8\,\textnormal{m}\cdot\textnormal{s}^{-2}$
is gravitational constant; and, $\mathbf{F}_{1}$ and $\mathbf{F}_{2}$
are 
\begin{align*}
\mathbf{F}_{1} & =-\mathbf{e}_{z}\pi\nu s\int_{0}^{\rho}\frac{\sigma_{iup}\rho_{1}d\rho_{1}}{\left(\rho_{1}^{2}+s^{2}\right)^{3/2}},
\end{align*}
 
\begin{align*}
\mathbf{F}_{2} & =\mathbf{e}_{z}\pi\nu\left(h-s\right)\int_{0}^{\rho}\frac{\sigma_{ilp}\rho_{2}d\rho_{2}}{\left[\rho_{2}^{2}+\left(h-s\right)^{2}\right]^{3/2}}.
\end{align*}
 Here, $\mathbf{F}_{1}$ is the force between the charged-particle
and the image charge induced by it on the surface of the anode. Similarly,
$\mathbf{F}_{2}$ is the force between the charged-particle and the
charge it induces at the cathode's surface. For the parallel plate
system which is microscopically large, but macroscopically small,
Eq.  (\ref{eq:FT-pre}) becomes\cite{Cho} 
\begin{align}
\mathbf{F}_{T}\left(s\right) & =\mathbf{e}_{z}\frac{\pi\epsilon_{0}\kappa_{3}\nu}{4}\left\{ \frac{\nu}{s^{2}}-\frac{\nu}{\left(h-s\right)^{2}}+\frac{\left[\gamma\left(b^{3}-a^{3}\right)-b^{3}\right]E_{p}}{s^{3}}\right.\nonumber \\
 & \left.+\frac{\left[\gamma\left(b^{3}-a^{3}\right)-b^{3}\right]E_{p}}{\left(h-s\right)^{3}}-8E_{p}\right\} -\mathbf{e}_{z}mg,\label{eq:FT-FINAL}
\end{align}
 where 
\[
m=\frac{4}{3}\pi\left[a^{3}\left(\rho_{m,1}-\rho_{m,2}\right)+b^{3}\rho_{m,2}\right]
\]
 with $\rho_{m,1}$ and $\rho_{m,2}$ representing mass densities
of the particle's core and shell, respectively.

\begin{figure}[h]
\begin{centering}
\includegraphics[width=0.9\columnwidth]{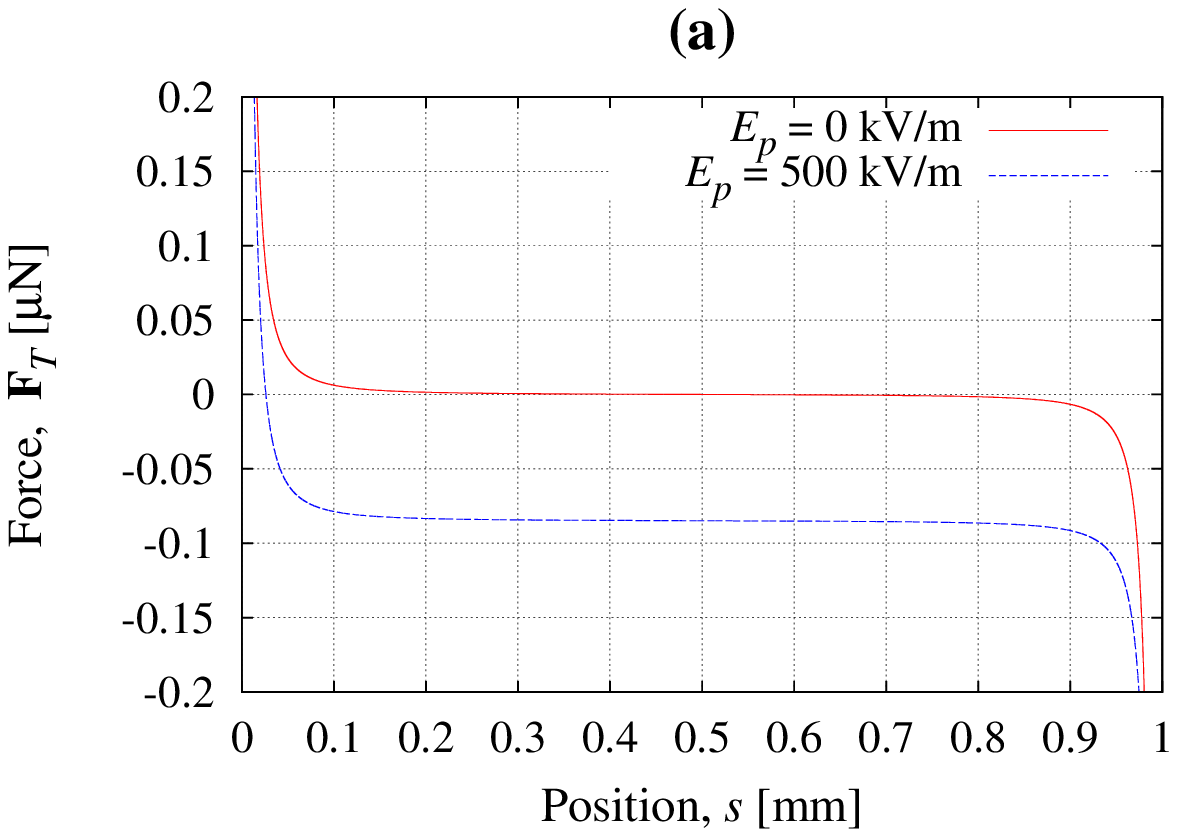}
\par\end{centering}

\begin{centering}
\includegraphics[width=0.9\columnwidth]{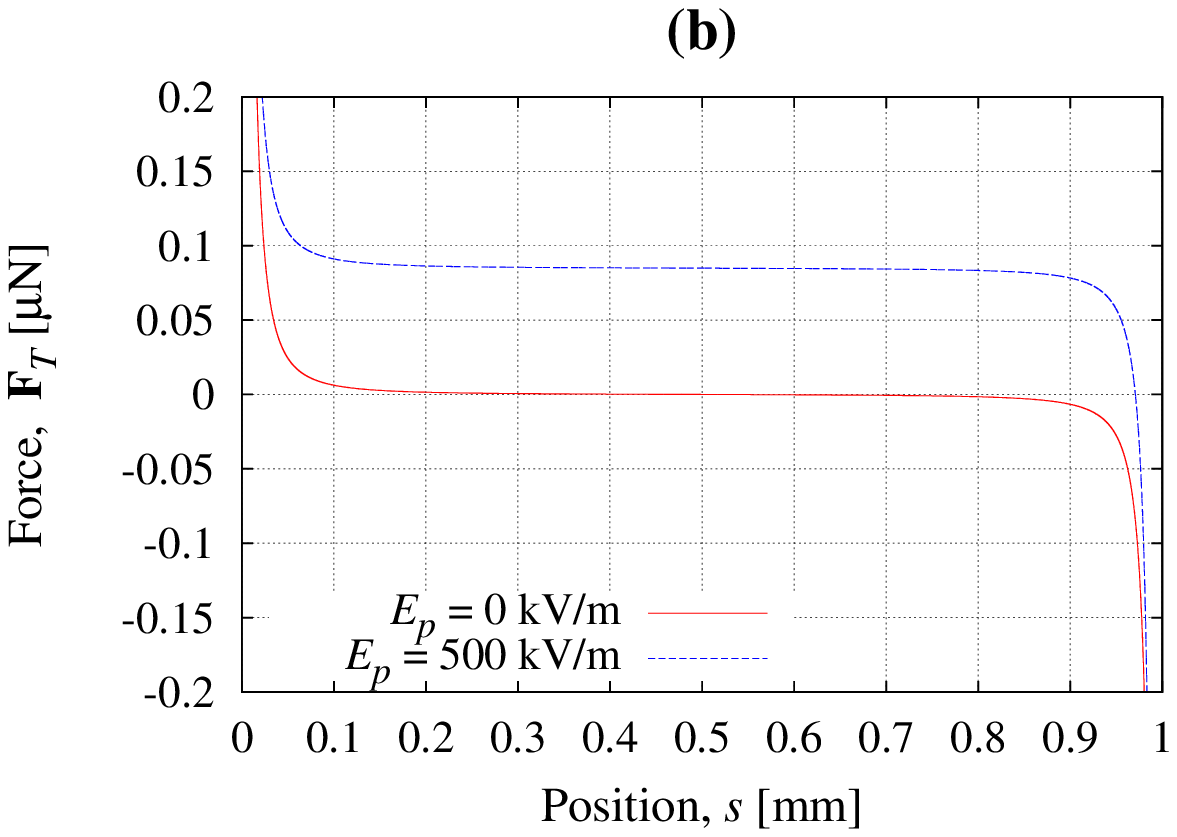}
\par\end{centering}

\caption{(Color online) Plot of $\mathbf{F}_{T}\left(s\right),$ Eq. (\ref{eq:FT-FINAL}),
for  $E_{p}=0\,\textnormal{kV}\cdot\textnormal{m}^{-1}$ and $E_{p}=500\,\textnormal{kV}\cdot\textnormal{m}^{-1}.$
(a) The case of a positively charged particle, where $\sigma_{1}=0.012\,\textnormal{C}\cdot\textnormal{m}^{-2}.$
(b) The case of a negatively charged particle, where $\sigma_{1}=-0.012\,\textnormal{C}\cdot\textnormal{m}^{-2}.$
In both (a) and (b), all other parameter values are as defined in
Eq. (\ref{eq:parameter-LQ}). The anode is located at the left side.
\label{fig:5}}
\end{figure}

The force $\mathbf{F}_{T}\left(s\right)$ of Eq. (\ref{eq:FT-FINAL})
is plotted using the following parameter values: 
\begin{equation}
\left\{ \begin{array}{c}
\kappa_{2}=6,\quad\kappa_{3}=1,\\
a=1.5\,\mu\textnormal{m},\quad h=1\,\textnormal{mm},\\
b-a=4\,\textnormal{nm},\\
V_{L}=0\,\textnormal{V},\\
\sigma_{2}=0\,\textnormal{C}\cdot\textnormal{m}^{-2}\textnormal{ (i.e., insulator)},\\
\rho_{m,1}=2700\,\textnormal{kg}\cdot\textnormal{m}^{-3},\\
\rho_{m,2}=3800\,\textnormal{kg}\cdot\textnormal{m}^{-3},
\end{array}\right.\label{eq:parameter-LQ}
\end{equation}
 where the choice of $4\,\textnormal{nm}$ for the shell's thickness
and a dielectric constant of $k_{2}\sim6$ are typical for alumina
nanoparticles.\cite{Japan-al2o3,al203} To compare the cases which
involve the positively and the negatively charged particles, the surface
charge densities of $\sigma_{1}=0.012\,\textnormal{C}\cdot\textnormal{m}^{-2}$
and $\sigma_{1}=-0.012\,\textnormal{C}\cdot\textnormal{m}^{-2}$ have
been considered. For the force dependence on the applied parallel
plate electric fields, $V_{T}=0\,\textnormal{V}$ and $V_{T}=500\,\textnormal{V}$
have been considered for comparison. The $V_{T}=0\,\textnormal{V}$
corresponds to the case where a charged-particle is placed between
the grounded plane-parallel plates. With $V_{L}=0\,\textnormal{V}$
and $h=1\,\textnormal{mm},$ the anode voltage of $V_{T}=500\,\textnormal{V}$
corresponds to the parallel plate electric field of $E_{p}=500\,\textnormal{kV}\cdot\textnormal{m}^{-1}.$
The results are shown in Fig. \ref{fig:5}(a). As expected, when both
plates are grounded, the charged-particle is attracted to the nearest
grounded plate. For instance, in the region where $s<0.5\,\textnormal{mm},$
the force $\mathbf{F}_{T}\left(s\right)$ is positive whereas, for
$s>0.5\,\textnormal{mm},$ the force becomes negative. Right at the
midway between the plates, i.e., $s=h/2,$ the net force on the charged-particle
vanishes. Why? This is because, at $s=h/2,$ the particle is equal
distance away from the surfaces of the anode and the cathode plates;
and, in such situation, the charges of opposite polarity are induced
at the surfaces of the anode and the cathode with equal magnitudes.
Finally, the presence of the parallel plate electric field, $\left\Vert \mathbf{E}_{p}\right\Vert >0,$
offsets the force to the negative axis for positively charged particles.
The resulting offset in the force gives rise to the oscillatory behavior
of the positively charged particle in vicinity of the anode. 

What happens to the force, $\mathbf{F}_{T}\left(s\right),$ when the
particle is negatively charged? Equation (\ref{eq:FT-FINAL}) has
been plotted using the same parameter values defined in Eq. (\ref{eq:parameter-LQ})
with the charge density of $\sigma_{1}=-0.012\,\textnormal{C}\cdot\textnormal{m}^{-2}$
for a negatively charged particle. The results are shown in Fig. \ref{fig:5}(b),
where $E_{p}=0\,\textnormal{kV}\cdot\textnormal{m}^{-1}$ and $E_{p}=500\,\textnormal{kV}\cdot\textnormal{m}^{-1}$
have been considered for comparison. As with the case of positively
charge particle, the negatively charged particle is attracted to the
closest electrode when $E_{p}=0\,\textnormal{kV}\cdot\textnormal{m}^{-1}.$
At the midway between the plates, i.e., $s=h/2,$ the negatively charged
particle feels no net force due to the fact that charges of opposite
polarity are induced at the facing surfaces of each plates with equal
magnitude. At the presence of the parallel plate electric field, i.e.,
$E_{p}=500\,\textnormal{kV}\cdot\textnormal{m}^{-1}$ in Fig. \ref{fig:5}(b),
the force offsets to the positive axis. The resulting offset in the
force gives rise to the oscillatory behavior of the negatively charged
particle in vicinity of the cathode. 

The physical charged-particle with structure cannot penetrate into
the surfaces of the anode and the cathode plates, of course. Therefore,
the parameter $s$ in Figs. \ref{fig:5}(a) and \ref{fig:5}(b) are
restricted to a domain $b\leq s\leq h-b,$ where $b$ is the charged-particle's
radius and $h$ is the gap between the anode and the cathode plates.
The case of $s=b$ corresponds to the situation where the particle
is in contact with the anode's surface; and, the case of $s=h-b$
corresponds to the situation where the particle is in contact with
the surface of the cathode. For this reason, the magnitude of the
force acting on the charged-particle remains finite at $s=b$ and
$s=h-b$ in Figs. \ref{fig:5}(a) and \ref{fig:5}(b).

\subsection{Positively charged particle with structure}

\subsubsection{Constituent forces}

It is worthwhile to discuss the constituent forces of $\mathbf{F}_{T}.$
When particle is sufficiently charged, the gravitational force becomes
negligible; and, Eq.  (\ref{eq:FT-FINAL}) can be expressed as 
\[
\mathbf{F}_{T}=\underbrace{\mathbf{f}_{1,1}+\mathbf{f}_{1,2}+\mathbf{f}_{1,3}}_{\mathbf{F}_{1}}+\underbrace{\mathbf{f}_{2,1}+\mathbf{f}_{2,2}+\mathbf{f}_{2,3}}_{\mathbf{F}_{2}},
\]
 where the constituent forces of $\mathbf{F}_{1}$ are 
\begin{align*}
\mathbf{f}_{1,1} & =\mathbf{e}_{z}\frac{\pi\epsilon_{0}\kappa_{3}\nu^{2}}{4s^{2}},\\
\mathbf{f}_{1,2} & =\mathbf{e}_{z}\frac{\pi\epsilon_{0}\kappa_{3}\nu\left[\gamma\left(b^{3}-a^{3}\right)-b^{3}\right]E_{p}}{4s^{3}},\\
\mathbf{f}_{1,3} & =-\mathbf{e}_{z}\pi\epsilon_{0}\kappa_{3}\nu E_{p};
\end{align*}
 and for $\mathbf{F}_{2},$ its constituent forces take on the form
given by 
\begin{align*}
\mathbf{f}_{2,1} & =-\mathbf{e}_{z}\frac{\pi\epsilon_{0}\kappa_{3}\nu^{2}}{4\left(h-s\right)^{2}},\\
\mathbf{f}_{2,2} & =\mathbf{e}_{z}\frac{\pi\epsilon_{0}\kappa_{3}\nu\left[\gamma\left(b^{3}-a^{3}\right)-b^{3}\right]E_{p}}{4\left(h-s\right)^{3}},\\
\mathbf{f}_{2,3} & =-\mathbf{e}_{z}\pi\epsilon_{0}\kappa_{3}\nu E_{p}.
\end{align*}
 Since $\nu>0$ and $1>\gamma>0,$ one finds 
\[
\gamma\left(b^{3}-a^{3}\right)-b^{3}<0;
\]
 and, the previous constituent force terms of $\mathbf{F}_{1}$ behave
as 
\[
\mathbf{f}_{1,1}\sim\mathbf{e}_{z}\frac{1}{s^{2}},\quad\mathbf{f}_{1,2}\sim-\mathbf{e}_{z}\frac{E_{p}}{s^{3}},\quad\mathbf{f}_{1,3}\sim-\mathbf{e}_{z}E_{p};
\]
 whereas for $\mathbf{F}_{2},$ 
\[
\mathbf{f}_{2,1}\sim-\mathbf{e}_{z}\frac{1}{\left(h-s\right)^{2}},\quad\mathbf{f}_{2,2}\sim-\mathbf{e}_{z}\frac{E_{p}}{\left(h-s\right)^{3}},\quad\mathbf{f}_{2,3}\sim-\mathbf{e}_{z}E_{p}.
\]
 Physically, $\mathbf{f}_{1,1}$ ($\mathbf{f}_{2,1}$) represents
an attractive force between charged particle and its image charge
at the surface of anode (cathode). The $\mathbf{f}_{1,3}$ ($\mathbf{f}_{2,3}$)
represents the usual force on charged-particle by parallel plate electric
field, $\mathbf{E}_{p}.$ The other force term, $\mathbf{f}_{1,2}$
($\mathbf{f}_{2,2}$), arises as a consequence of a structured particle
that becomes polarized under $\mathbf{E}_{p}.$ Such force vanishes
in the absence $\mathbf{E}_{p}.$ 

\begin{figure}[h]
\begin{centering}
\includegraphics[width=0.9\columnwidth]{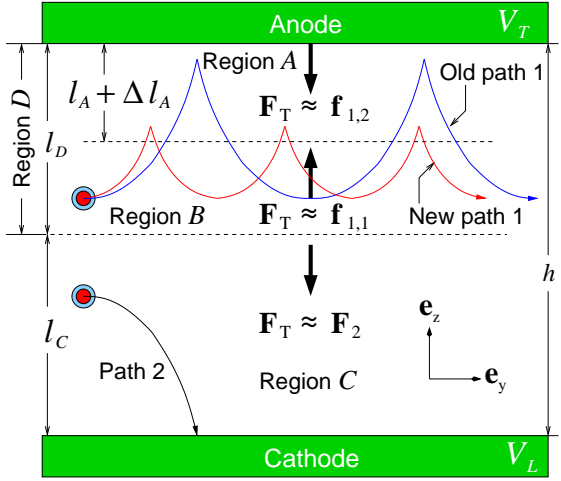}
\par\end{centering}

\caption{(Color online) Schematic illustration of positively charged, structured,
particle oscillating in vicinity of the anode. The dominant constituent
force terms are shown in regions $A$ and $B.$ No oscillation mode
exists near the cathode, region $C,$ for a positive particle. The
$\textnormal{old path 1,}$ $\textnormal{new path 1,}$ and the $\textnormal{path 2}$
represent the schematic plot of $z_{d}\left(t\right)$ versus time,
where time is the horizontal axis. The $\textnormal{new path 1}$
corresponds to a case where $E_{p}$ is higher than the one in $\textnormal{old path 1.}$
\label{fig:6}}
\end{figure}

So, what gives rise to charged-particle oscillation? The $\mathbf{F}_{2}$
cannot generate oscillations because $\mathbf{f}_{2,1},$ $\mathbf{f}_{2,2},$
and $\mathbf{f}_{2,3}$ are all directed in the same direction. The
$\mathbf{F}_{1},$ on the other hand, contains constituent forces
with opposite directions that compete one another; and, such terms
generate oscillations. For instance, at distances very close to the
anode, $\mathbf{F}_{1}\approx\mathbf{f}_{1,2}\sim-\mathbf{e}_{z}E_{p}s^{-3};$
and, such force is responsible for the Coulomb repulsion in region
$A$ of Fig. \ref{fig:3}(a). This force decays rapidly with distance.
Consequently, outside of the region $A,$ the contributions from $\mathbf{f}_{1,2}$
become negligible. In region $B,$ e.g., Fig. \ref{fig:3}(b), the
force on the particle is dominated by $\mathbf{F}_{1}\approx\mathbf{f}_{1,1}\sim\mathbf{e}_{z}s^{-2}.$
This force attracts the charged-particle back to the anode. It is
this ``push-pull'' competition between $\mathbf{f}_{1,1}$ and $\mathbf{f}_{1,2}$
that gives rise to charged-particle oscillation in region $D,$ as
illustrated in Fig. \ref{fig:6}. 

The width $l_{A}$ of region $A$ increases with $E_{p}$ as a consequence
of $\mathbf{f}_{1,2}\sim-\mathbf{e}_{z}E_{p}s^{-3}.$ If region $A$
widens by $\Delta l_{A},$ the width of region $B$ decreases by the
same amount. This is because the width $l_{D}$ of region $D$ remains
fixed for a given charged-particle of constant excess charge $q.$
Consequently, the particle's oscillation frequency increases with
$E_{p}$ in region $D.$ Such is schematically illustrated in Fig.
\ref{fig:6}. The $\textnormal{new path 1},$ which is the plot of
$z_{d}\left(t\right)$ versus time corresponding to the case of higher
$E_{p},$ has a higher oscillation frequency than the $\textnormal{old path 1}.$

\subsubsection{Potential energy}

\begin{figure}[h]
\begin{centering}
\includegraphics[width=1\columnwidth]{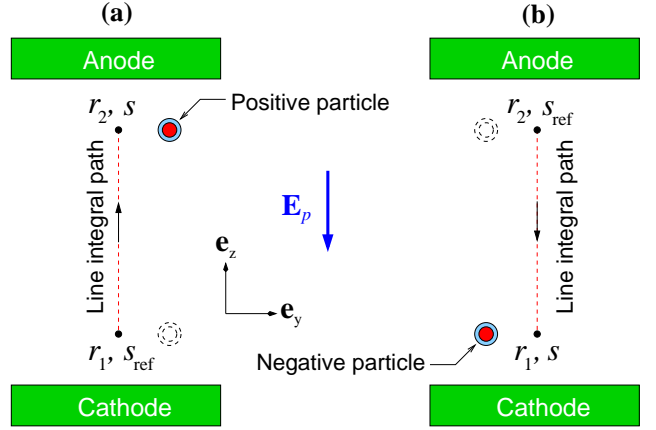}
\par\end{centering}

\caption{(Color online) Illustrates the integral path in a line integral for
a given parallel plate electric field $\mathbf{E}_{p}.$ (a) The case
of positively charged particle. (b) The case of negatively charged
particle. \label{fig:7}}
\end{figure}

By definition, the force $\mathbf{F}$ is defined as the negative
gradient of the potential energy function $U,$ 
\[
\mathbf{F}\equiv-\nabla U.
\]
 For a one dimensional force, such as the one in Eq. (\ref{eq:FT-FINAL}),
this implies 
\[
\mathbf{F}_{T}\left(s\right)=-\mathbf{e}_{z}\frac{dU}{ds},
\]
 where $U\equiv U\left(s\right).$ The one dimensional potential energy
function, $U\left(s\right),$ becomes 
\begin{equation}
U\left(s\right)=-\int_{s_{\textnormal{ref}}}^{s}\mathbf{F}_{T}\cdot\mathbf{e}_{z}ds,\label{eq:path-integral}
\end{equation}
 where $s_{\textnormal{ref}}$ is the reference point in which $U\left(s_{\textnormal{ref}}\right)<U\left(s\right).$
For instance, at the presence of the parallel plate electric field,
$\mathbf{E}_{p},$ the potential energy of a positively charged particle
may be computed by integrating the line integral along the path illustrated
in Fig. \ref{fig:7}(a). On the other hand, the potential energy for
a negatively charged particle subjected to the same parallel plate
electric field can be computed by integrating the line integral along
the path illustrated in Fig. \ref{fig:7}(b). Consequently, the presence
of electric field, $\mathbf{E}_{p},$ as well as its directions and
the polarity of the charged-particle affect the choice of $s_{\textnormal{ref}}$
in the line integral of Eq. (\ref{eq:path-integral}). For such reason,
I shall only consider the case where $\left\Vert \mathbf{E}_{p}\right\Vert >0$
for the evaluation of $U\left(s\right).$ 

That said, Eq. (\ref{eq:FT-FINAL}) is inserted for $\mathbf{F}_{T}$
in Eq. (\ref{eq:path-integral}) to yield 
\begin{align}
U\left(s\right) & =-\frac{\pi\epsilon_{0}\kappa_{3}\nu}{4}\int_{s_{\textnormal{ref}}}^{s}\left\{ \frac{\nu}{s^{2}}-\frac{\nu}{\left(h-s\right)^{2}}\right.\nonumber \\
 & +\frac{\left[\gamma\left(b^{3}-a^{3}\right)-b^{3}\right]E_{p}}{s^{3}}\nonumber \\
 & \left.+\frac{\left[\gamma\left(b^{3}-a^{3}\right)-b^{3}\right]E_{p}}{\left(h-s\right)^{3}}-8E_{p}\right\} ds\nonumber \\
 & +mg\int_{s_{\textnormal{ref}}}^{s}ds,\label{eq:U(s)-pre}
\end{align}
 where $E_{p}>0.$ Equation (\ref{eq:U(s)-pre}) is evaluated utilizing
the following integral formulas: 
\begin{align*}
\int_{s_{\textnormal{ref}}}^{s}ds & =s-s_{\textnormal{ref}},
\end{align*}
 
\begin{align*}
\int_{s_{\textnormal{ref}}}^{s}\frac{1}{s^{2}}ds & =\frac{1}{s_{\textnormal{ref}}}-\frac{1}{s},
\end{align*}
 
\begin{align*}
\int_{s_{\textnormal{ref}}}^{s}\frac{1}{s^{3}}ds & =\frac{1}{2s_{\textnormal{ref}}^{2}}-\frac{1}{2s^{2}},
\end{align*}
 
\begin{align*}
\int_{s_{\textnormal{ref}}}^{s}\frac{1}{\left(h-s\right)^{2}}ds & =\frac{1}{h-s}-\frac{1}{h-s_{\textnormal{ref}}},
\end{align*}
 
\begin{align*}
\int_{s_{\textnormal{ref}}}^{s}\frac{1}{\left(h-s\right)^{3}}ds & =\frac{1}{2\left(h-s\right)^{2}}-\frac{1}{2\left(h-s_{\textnormal{ref}}\right)^{2}}.
\end{align*}
 The result is 
\begin{align}
U\left(s\right) & =\frac{\pi\epsilon_{0}\kappa_{3}\nu}{4}\left\{ \frac{\nu}{s}+\frac{\nu}{h-s}+\frac{\left[\gamma\left(b^{3}-a^{3}\right)-b^{3}\right]E_{p}}{2s^{2}}\right.\nonumber \\
 & \left.-\frac{\left[\gamma\left(b^{3}-a^{3}\right)-b^{3}\right]E_{p}}{2\left(h-s\right)^{2}}+8E_{p}s\right\} +mgs\nonumber \\
 & -\frac{\pi\epsilon_{0}\kappa_{3}\nu}{4}\left\{ \frac{\nu}{s_{\textnormal{ref}}}+\frac{\nu}{h-s_{\textnormal{ref}}}+\frac{\left[\gamma\left(b^{3}-a^{3}\right)-b^{3}\right]E_{p}}{2s_{\textnormal{ref}}^{2}}\right.\nonumber \\
 & \left.-\frac{\left[\gamma\left(b^{3}-a^{3}\right)-b^{3}\right]E_{p}}{2\left(h-s_{\textnormal{ref}}\right)^{2}}+8E_{p}s_{\textnormal{ref}}\right\} -mgs_{\textnormal{ref}},\label{eq:U(s)-pre-1}
\end{align}
 where $E_{p}>0.$ The parameter $s$ is related to the parameter
$z_{d},$ which is defined in Fig. \ref{fig:4}, by 
\begin{equation}
s=z_{d}+b\quad\textnormal{and}\quad\ddot{s}=\ddot{z}_{d}.\label{eq:s-and-z-relation}
\end{equation}
 Utilizing this definition, Eq. (\ref{eq:U(s)-pre-1}) can be rewritten
as 
\begin{align*}
U\left(z_{d}\right) & =\frac{\pi\epsilon_{0}\kappa_{3}\nu}{4}\left\{ \frac{\nu}{z_{d}+b}+\frac{\nu}{h-z_{d}-b}\right.\\
 & +\frac{\left[\gamma\left(b^{3}-a^{3}\right)-b^{3}\right]E_{p}}{2\left(z_{d}+b\right)^{2}}-\frac{\left[\gamma\left(b^{3}-a^{3}\right)-b^{3}\right]E_{p}}{2\left(h-z_{d}-b\right)^{2}}\\
 & \left.+8E_{p}\left(z_{d}+b\right)\vphantom{\frac{\nu}{z_{d}+b}}\right\} +mg\left(z_{d}+b\right)\\
 & -\frac{\pi\epsilon_{0}\kappa_{3}\nu}{4}\left\{ \frac{\nu}{s_{\textnormal{ref}}}+\frac{\nu}{h-s_{\textnormal{ref}}}+\frac{\left[\gamma\left(b^{3}-a^{3}\right)-b^{3}\right]E_{p}}{2s_{\textnormal{ref}}^{2}}\right.\\
 & \left.-\frac{\left[\gamma\left(b^{3}-a^{3}\right)-b^{3}\right]E_{p}}{2\left(h-s_{\textnormal{ref}}\right)^{2}}+8E_{p}s_{0}\right\} -mgs_{\textnormal{ref}},
\end{align*}
 where $E_{p}>0.$ I shall set $s_{\textnormal{ref}}$ at the midway
between the parallel plates, 
\begin{equation}
s_{\textnormal{ref}}=\frac{h}{2};\label{eq:sref-pos}
\end{equation}
 and the $U\left(z_{d}\right)$ becomes 
\begin{align}
U\left(z_{d}\right) & =\frac{\pi\epsilon_{0}\kappa_{3}\nu}{4}\left\{ \frac{\nu}{z_{d}+b}+\frac{\nu}{h-z_{d}-b}\right.\nonumber \\
 & +\frac{\left[\gamma\left(b^{3}-a^{3}\right)-b^{3}\right]E_{p}}{2\left(z_{d}+b\right)^{2}}-\frac{\left[\gamma\left(b^{3}-a^{3}\right)-b^{3}\right]E_{p}}{2\left(h-z_{d}-b\right)^{2}}\nonumber \\
 & \left.+8E_{p}\left(z_{d}+b\right)\vphantom{\frac{\nu}{z_{d}+b}}\right\} +mg\left(z_{d}+b\right)\nonumber \\
 & -\frac{\pi\epsilon_{0}\kappa_{3}\nu}{h}\left(\nu+E_{p}h^{2}\right)-\frac{1}{2}mgh,\label{eq:U(zd)-FINAL}
\end{align}
 where $E_{p}>0;$ and, the explicit expressions for $E_{p},$ $\gamma,$
and $\nu$ are defined in Eqs. (\ref{eq:Ep-DEF}) and (\ref{eq:alpla-beta-gama-lambda-nu}):
\begin{align*}
E_{p} & =\frac{V_{T}-V_{L}}{h},\quad V_{T}>V_{L};\\
\gamma & =\frac{3\kappa_{3}b^{3}}{\left(\kappa_{2}+2\kappa_{3}\right)b^{3}+2\left(\kappa_{2}-\kappa_{3}\right)a^{3}},\\
\nu & =\frac{2a\left(b-a\right)\sigma_{1}}{\epsilon_{0}\kappa_{2}}+\frac{a^{2}\sigma_{1}+b^{2}\sigma_{2}}{\epsilon_{0}\kappa_{3}}.
\end{align*}

\begin{figure}[h]
\begin{centering}
\includegraphics[width=0.85\columnwidth]{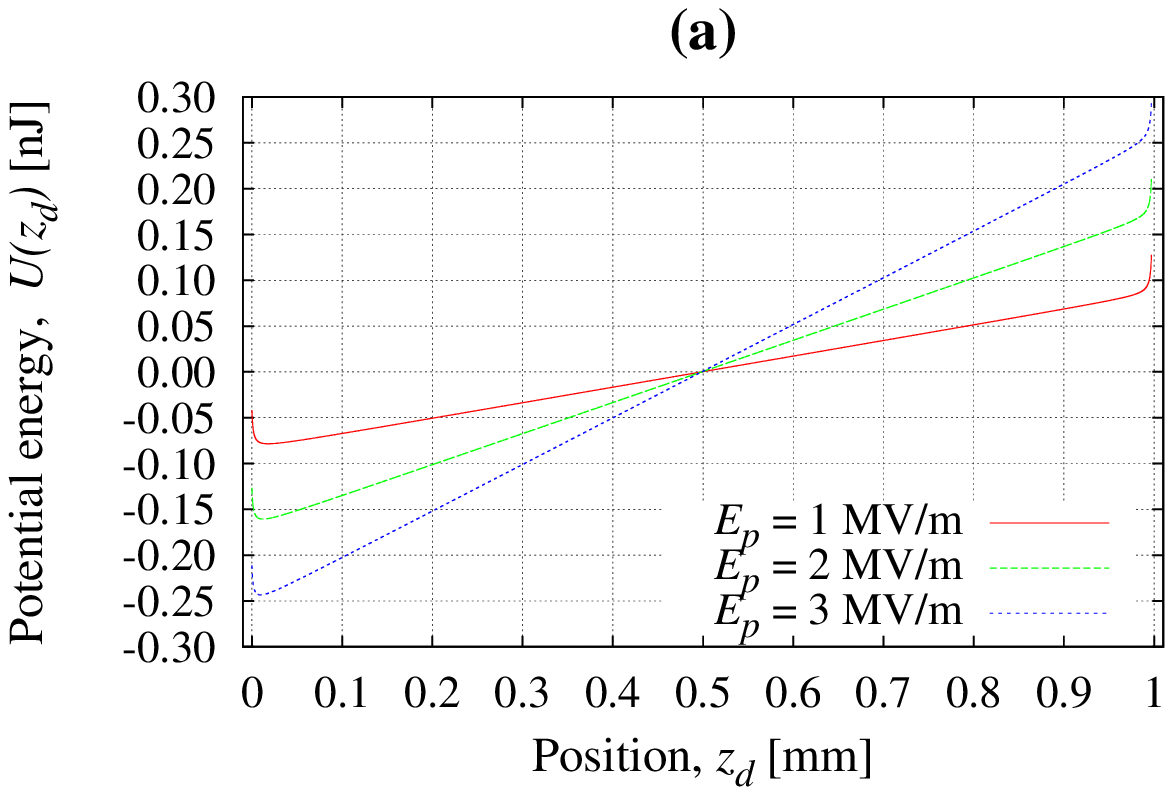}
\par\end{centering}

\begin{centering}
\includegraphics[width=0.85\columnwidth]{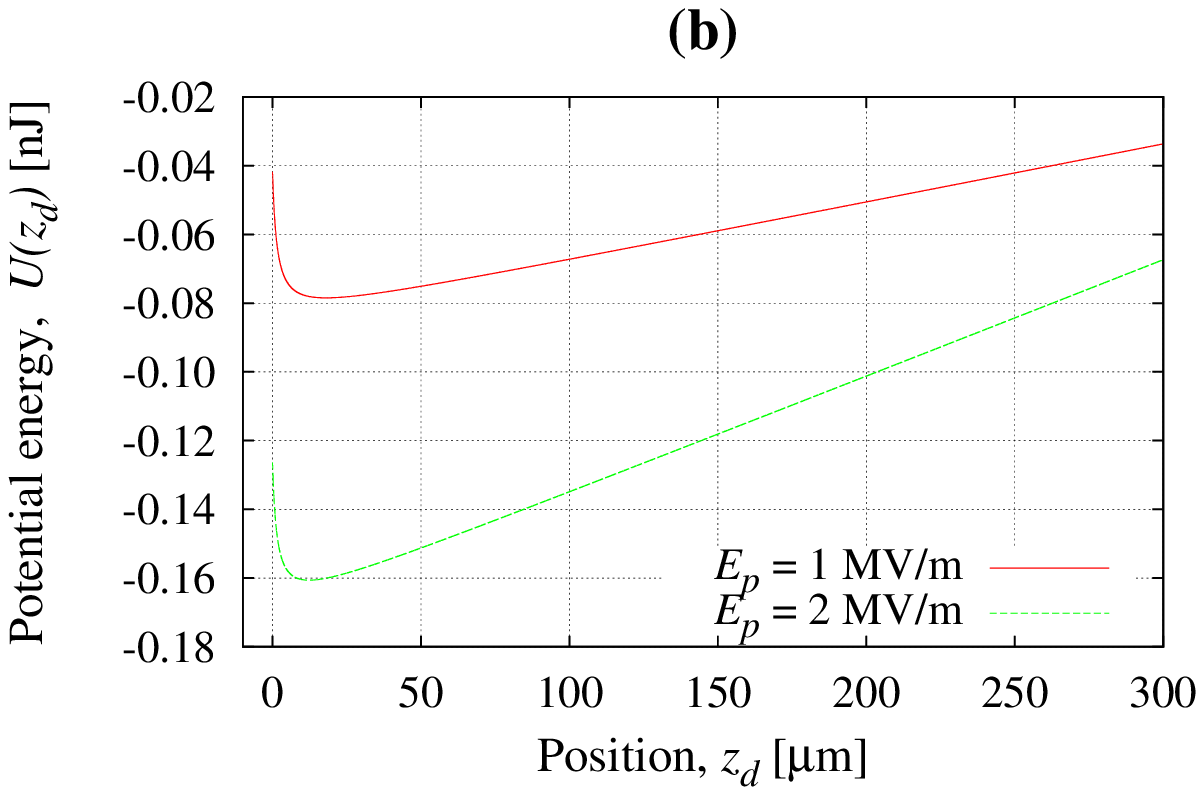}
\par\end{centering}

\caption{(Color online) (a) Plot of the potential energy function, $U\left(z_{d}\right)$
of Eq. (\ref{eq:U(zd)-FINAL}), for $E_{p}=1\,\textnormal{MV}\cdot\textnormal{m}^{-1},$
$E_{p}=2\,\textnormal{MV}\cdot\textnormal{m}^{-1},$ and $E_{p}=3\,\textnormal{MV}\cdot\textnormal{m}^{-1}.$
The particle has a charge density of $\sigma_{1}=0.012\,\textnormal{C}\cdot\textnormal{m}^{-2}$
and all other parameter values are as defined in Eq. (\ref{eq:parameter-LQ}).
(b) Enlarged plot of $U\left(z_{d}\right)$ for domain $0<z_{d}<h/2,$
where $h=1\,\textnormal{mm}.$  In the plot, the anode is located
at $z_{d}=0\,\mu\textnormal{m}.$ \label{fig:8}}
\end{figure}

Equation (\ref{eq:U(zd)-FINAL}) is plotted for a positively charged
particle in which $\sigma_{1}=0.012\,\textnormal{C}\cdot\textnormal{m}^{-2}$
and all other parameter values are same as defined in Eq. (\ref{eq:parameter-LQ}),
\[
\left\{ \begin{array}{c}
\kappa_{2}=6,\quad\kappa_{3}=1,\\
a=1.5\,\mu\textnormal{m},\quad h=1\,\textnormal{mm},\\
b-a=4\,\textnormal{nm},\\
V_{L}=0\,\textnormal{V},\\
\sigma_{2}=0\,\textnormal{C}\cdot\textnormal{m}^{-2}\textnormal{ (i.e., insulator)},\\
\rho_{m,1}=2700\,\textnormal{kg}\cdot\textnormal{m}^{-3},\\
\rho_{m,2}=3800\,\textnormal{kg}\cdot\textnormal{m}^{-3}.
\end{array}\right.
\]
 For the plot, the anode voltages of $V_{T}=1\,\textnormal{kV},$
$V_{T}=2\,\textnormal{kV},$ and $V_{T}=3\,\textnormal{kV}$ are considered
for comparison. For $V_{L}=0\,\textnormal{V}$ and $h=1\,\textnormal{mm},$
these anode voltages correspond to the parallel plate electric field
strengths of $E_{p}=1\,\textnormal{MV}\cdot\textnormal{m}^{-1},$
$E_{p}=2\,\textnormal{MV}\cdot\textnormal{m}^{-1},$ and $E_{p}=3\,\textnormal{MV}\cdot\textnormal{m}^{-1},$
respectively. The results are shown in Figs. \ref{fig:8}(a) and \ref{fig:8}(b),
where the ``potential well'' corresponding to each $E_{p}$ are
formed near the anode. The depth of the ``potential well,'' wherein
the charged-particle can have oscillatory solutions, depends on the
magnitude of the applied parallel plate electric field. Because the
physical particle cannot penetrate into the surface of the anode,
the parameter $z_{d}$ cannot be negative valued. When $z_{d}=0\,\mu\textnormal{m},$
the particle is right on the anode's surface; and, this restricts
the height of the potential well at $z_{d}=0\,\mu\textnormal{m}$
to a finite value, which can be verified from Fig. \ref{fig:8}(b).
In the case of $E_{p}=1\,\textnormal{MV}\cdot\textnormal{m}^{-1},$
the width of the potential well is approximately $\sim250\,\mu\textnormal{m}$
and the particle is restricted to $0\,\mu\textnormal{m}<z_{d}\lesssim250\,\mu\textnormal{m}$
for oscillations. The width of the potential well decreases with the
applied parallel plate electric field. For instance, in the case of
$E_{p}=2\,\textnormal{MV}\cdot\textnormal{m}^{-1},$ the potential
well width is approximately $\sim125\,\mu\textnormal{m}$ and the
particle is restricted to $0\,\mu\textnormal{m}<z_{d}\lesssim125\,\mu\textnormal{m}$
for oscillations. Physically, this corresponds to the narrowing of
the positive glow region with increased $E_{p}.$ 

\subsubsection{Dynamics in nonrelativistic regime }

\begin{figure}[h]
\begin{centering}
\includegraphics[width=0.9\columnwidth]{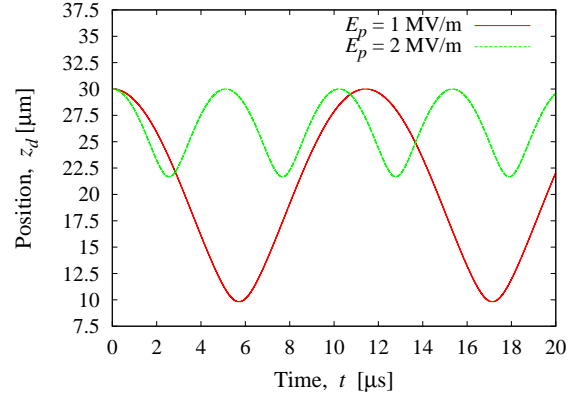}
\par\end{centering}

\caption{(Color online) Oscillating positively charged nonrelativistic particle
with structure in vicinity of the anode. The plot of $z_{d}\left(t\right)$
has been obtained from Eq. (\ref{eq:ODE-newton}) using the initial
conditions specified in Eq. (\ref{eq:IC-positive}) and the parameter
values specified in Eq. (\ref{eq:parameter-LQ}) with $\sigma_{1}=0.012\,\textnormal{C}\cdot\textnormal{m}^{-2}.$
In the plot, the anode is located at $z_{d}=0\,\mu\textnormal{m}.$
\label{fig:9}}
\end{figure}

The particle's equation of motion, in the nonrelativistic limit, is
obtained by solving 
\begin{equation}
\mathbf{e}_{z}m\ddot{s}=\mathbf{F}_{T},\label{eq:newton-law}
\end{equation}
 where $\ddot{s}\equiv d^{2}s/dt^{2}$ is the particle's acceleration.
Insertion of Eq.  (\ref{eq:FT-FINAL}) for $\mathbf{F}_{T}$ in Eq.
(\ref{eq:newton-law}) yields 
\begin{align*}
\ddot{s} & =\frac{\pi\epsilon_{0}\kappa_{3}\nu}{4m}\left\{ \frac{\nu}{s^{2}}-\frac{\nu}{\left(h-s\right)^{2}}+\frac{\left[\gamma\left(b^{3}-a^{3}\right)-b^{3}\right]E_{p}}{s^{3}}\right.\\
 & \left.+\frac{\left[\gamma\left(b^{3}-a^{3}\right)-b^{3}\right]E_{p}}{\left(h-s\right)^{3}}-8E_{p}\right\} -g,
\end{align*}
 where $\mathbf{e}_{z}$ has been dropped for convenience. Utilizing
the relations in Eq. (\ref{eq:s-and-z-relation}), this can be rewritten
as 
\begin{align}
\ddot{z}_{d} & =\frac{\pi\epsilon_{0}\kappa_{3}\nu}{4m}\left\{ \frac{\nu}{\left(z_{d}+b\right)^{2}}-\frac{\nu}{\left(h-z_{d}-b\right)^{2}}\right.\nonumber \\
 & +\frac{\left[\gamma\left(b^{3}-a^{3}\right)-b^{3}\right]E_{p}}{\left(z_{d}+b\right)^{3}}\nonumber \\
 & \left.+\frac{\left[\gamma\left(b^{3}-a^{3}\right)-b^{3}\right]E_{p}}{\left(h-z_{d}-b\right)^{3}}-8E_{p}\right\} -g.\label{eq:ODE-newton}
\end{align}
 Equation (\ref{eq:ODE-newton}) is solved via Runge-Kutta method.
For the parameter values, I shall use the same values specified in
Eq. (\ref{eq:parameter-LQ}) with $\sigma_{1}=0.012\,\textnormal{C}\cdot\textnormal{m}^{-2}.$
For the initial conditions, I shall choose 
\begin{equation}
z_{d}\left(0\right)=30\,\mu\textnormal{m}\quad\textnormal{and}\quad\dot{z}_{d}\left(0\right)=0.\label{eq:IC-positive}
\end{equation}
 The initial condition for the particle's position has been chosen
from the consideration of the potential energy function illustrated
in Fig. \ref{fig:8}; for instance, $z_{d}\left(0\right)=30\,\mu\textnormal{m}$
is within the potential well illustrated in Fig. \ref{fig:8}(b).
The anode voltages of $V_{T}=1\,\textnormal{kV}$ and $V_{T}=2\,\textnormal{kV}$
have been considered for comparison. For $V_{L}=0\,\textnormal{V}$
and $h=1\,\textnormal{mm},$ the anode voltages of $V_{T}=1\,\textnormal{kV}$
and $V_{T}=2\,\textnormal{kV}$ corresponds to parallel plate electric
field strengths of $E_{p}=1\,\textnormal{MV}\cdot\textnormal{m}^{-1}$
and $E_{p}=2\,\textnormal{MV}\cdot\textnormal{m}^{-1},$ respectively.
The results are shown in Fig. \ref{fig:9}. The dependence of oscillation
frequency on $E_{p}$ is consistent with the argument discussed previously
in Fig. \ref{fig:6}. 

\subsubsection{Dynamics in relativistic regime }

In the relativistic generalization, the equation of motion for the
charged-particle is obtained by solving 
\begin{equation}
\mathbf{e}_{z}\frac{d}{dt}\left[\frac{m\dot{s}}{\sqrt{1-\left(\frac{\dot{s}}{c}\right)^{2}}}\right]=\mathbf{F}_{T},\label{eq:einstein}
\end{equation}
 where $c=3\times10^{8}\,\textnormal{m}\cdot\textnormal{s}^{-1}$
is the speed of light in vacuum and $\dot{s}=ds/dt$ is the charged-particle's
speed. Insertion of Eq.  (\ref{eq:FT-FINAL}) for $\mathbf{F}_{T}$
in Eq. (\ref{eq:einstein}), and after some rearrangements, yields\cite{Cho}
\begin{align}
\ddot{s} & =\left(1-\frac{\dot{s}^{2}}{c^{2}}\right)^{3/2}\left(\frac{\pi\epsilon_{0}\kappa_{3}\nu}{4m}\left\{ \frac{\nu}{s^{2}}-\frac{\nu}{\left(h-s\right)^{2}}\right.\right.\nonumber \\
 & +\frac{\left[\gamma\left(b^{3}-a^{3}\right)-b^{3}\right]E_{p}}{s^{3}}\nonumber \\
 & \left.\left.+\frac{\left[\gamma\left(b^{3}-a^{3}\right)-b^{3}\right]E_{p}}{\left(h-s\right)^{3}}-8E_{p}\right\} -g\right),\label{eq:ODE-einstein-pre}
\end{align}
 where $\mathbf{e}_{z}$ has been dropped for convenience. The $z_{d}$
parameter defined in Fig. \ref{fig:4} is related to the parameter
$s$ by 
\[
s=z_{d}+b,\quad\dot{s}=\dot{z}_{d},\quad\ddot{s}=\ddot{z}_{d};
\]
 and the expression in Eq.  (\ref{eq:ODE-einstein-pre}) becomes 
\begin{align}
\ddot{z}_{d} & =\left(1-\frac{\dot{z}_{d}^{2}}{c^{2}}\right)^{3/2}\left(\frac{\pi\epsilon_{0}\kappa_{3}\nu}{4m}\left\{ \frac{\nu}{\left(z_{d}+b\right)^{2}}\right.\right.\nonumber \\
 & -\frac{\nu}{\left(h-z_{d}-b\right)^{2}}+\frac{\left[\gamma\left(b^{3}-a^{3}\right)-b^{3}\right]E_{p}}{\left(z_{d}+b\right)^{3}}\nonumber \\
 & \left.\left.+\frac{\left[\gamma\left(b^{3}-a^{3}\right)-b^{3}\right]E_{p}}{\left(h-z_{d}-b\right)^{3}}-8E_{p}\right\} -g\right).\label{eq:ODE-einstein-fin}
\end{align}
 Equation (\ref{eq:ODE-einstein-fin}) describes the charged-particle's
motion at all speed ranges.

\subsection{Negatively charged particle with structure}

\subsubsection{Constituent forces}

\begin{figure}[h]
\begin{centering}
\includegraphics[width=0.9\columnwidth]{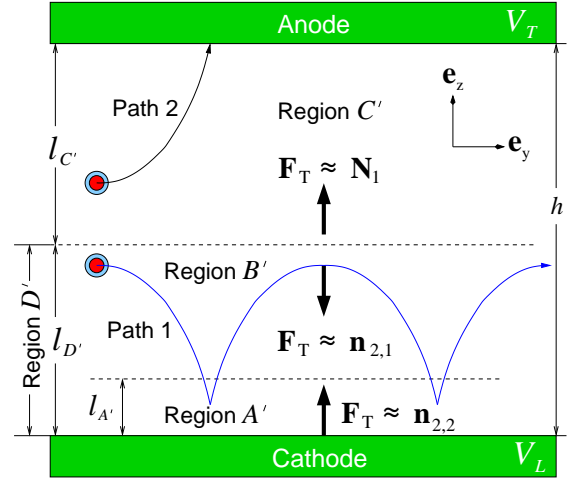}
\par\end{centering}

\caption{(Color online) Illustration of negatively charged, structured, particle
oscillating in vicinity of the cathode, i.e., the regions $A'$ and
$B'.$ The dominant constituent force terms are shown in regions $A'$
and $B'.$ No oscillation mode exists near the anode, i.e., the region
$C'.$ The width of the region $A'$ is identified by $l_{A'},$ and
the width of the region $B'$ is given by $l_{D'}-l_{A'},$ where
$l_{D'}$ is the borderline between the regions $B'$ and $C'.$ The
$\textnormal{path 1}$ and $\textnormal{path 2}$ represent the schematic
plot of $z_{d}\left(t\right)$ versus time, where the horizontal axis
is the time. \label{fig:10}}
\end{figure}

For negatively charged core-shell structured particle, neglecting
the gravity, $\mathbf{F}_{T}$ of Eq.  (\ref{eq:FT-FINAL}) gets modified
as\cite{Cho} 
\[
\mathbf{F}_{T}=\underbrace{\mathbf{n}_{1,1}+\mathbf{n}_{1,2}+\mathbf{n}_{1,3}}_{\mathbf{N}_{1}}+\underbrace{\mathbf{n}_{2,1}+\mathbf{n}_{2,2}+\mathbf{n}_{2,3}}_{\mathbf{N}_{2}},
\]
 where the constituent forces of $\mathbf{N}_{1}$ are 
\begin{align*}
\mathbf{n}_{1,1} & =\mathbf{e}_{z}\frac{\pi\epsilon_{0}\kappa_{3}\nu^{2}}{4s^{2}}\sim\mathbf{e}_{z}\frac{1}{s^{2}},\\
\mathbf{n}_{1,2} & =\mathbf{e}_{z}\frac{\pi\epsilon_{0}\kappa_{3}\left|\nu\right|\left|\gamma\left(b^{3}-a^{3}\right)-b^{3}\right|E_{p}}{4s^{3}}\sim\mathbf{e}_{z}\frac{E_{p}}{s^{3}},\\
\mathbf{n}_{1,3} & =\mathbf{e}_{z}\pi\epsilon_{0}\kappa_{3}\left|\nu\right|E_{p}\sim\mathbf{e}_{z}E_{p};
\end{align*}
 and the constituent forces of $\mathbf{N}_{2}$ are given by 
\begin{align*}
\mathbf{n}_{2,1} & =-\mathbf{e}_{z}\frac{\pi\epsilon_{0}\kappa_{3}\nu^{2}}{4\left(h-s\right)^{2}}\sim-\mathbf{e}_{z}\frac{1}{\left(h-s\right)^{2}},\\
\mathbf{n}_{2,2} & =\mathbf{e}_{z}\frac{\pi\epsilon_{0}\kappa_{3}\left|\nu\right|\left|\gamma\left(b^{3}-a^{3}\right)-b^{3}\right|E_{p}}{4\left(h-s\right)^{3}}\sim\mathbf{e}_{z}\frac{E_{p}}{\left(h-s\right)^{3}},\\
\mathbf{n}_{2,3} & =\mathbf{e}_{z}\pi\epsilon_{0}\kappa_{3}\left|\nu\right|E_{p}\sim\mathbf{e}_{z}E_{p}.
\end{align*}
 Here, $\mathbf{N}_{1}$ is the force between charged-particle and
its image charge formed at the anode's surface whereas $\mathbf{N}_{2}$
corresponds to the force between charged-particle and its image charge
contribution at the cathode's surface. The notations $\left(\mathbf{N}_{1},\mathbf{N}_{2}\right)$
are introduced to distinguish from $\left(\mathbf{F}_{1},\mathbf{F}_{2}\right)$
of the positive charged-particle case. In the case of negative charged-particle,
the force $\mathbf{N}_{2}$ gives rise to oscillations; and, the particle
oscillates in vicinity of the cathode, as schematically illustrated
in Fig. \ref{fig:10}. 

\begin{figure}[h]
\begin{centering}
\includegraphics[width=0.9\columnwidth]{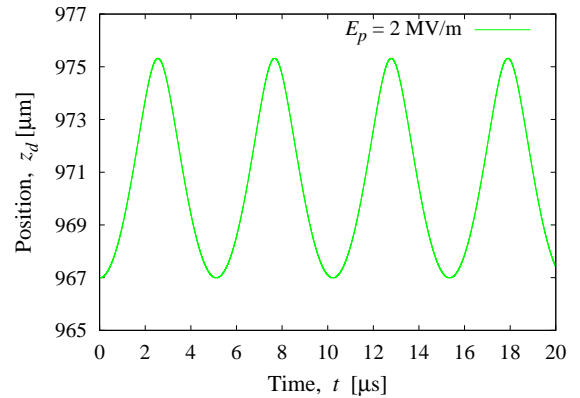}
\par\end{centering}

\caption{(Color online) Oscillating negatively charged nonrelativistic particle
with structure in vicinity of the cathode. The $z_{d}\left(t\right)$
of Eq. (\ref{eq:ODE-newton}) has been plotted using the initial value
conditions defined in Eq. (\ref{eq:IC-negative}) and the parameter
values defined in Eq. (\ref{eq:parameter-LQ}) with $\sigma_{1}=-0.012\,\textnormal{C}\cdot\textnormal{m}^{-2}$
and $V_{T}=2\,\textnormal{kV}$ (or $E_{p}=2\,\textnormal{MV}\cdot\textnormal{m}^{-1}$).
The anode plate is located at $z_{d}=0\,\mu\textnormal{m}.$ \label{fig:11}}
\end{figure}

To validate the argument illustrated in Fig. \ref{fig:10}, Eq.  (\ref{eq:ODE-newton})
is evaluated for a negatively charged particle, $\sigma_{1}=-0.012\,\textnormal{C}\cdot\textnormal{m}^{-2},$
using the following initial conditions: 
\begin{equation}
z_{d}\left(0\right)=h-2b-z_{d}^{\prime}\quad\textnormal{and}\quad\dot{z}_{d}\left(0\right)=0,\label{eq:IC-negative}
\end{equation}
 where $z_{d}^{\prime}=30\,\mu\textnormal{m}.$ To compare the result
against the positively charged particle situation discussed in Fig.
\ref{fig:9}, the particle's initial position has been assigned such
that there is a gap of $30\,\mu\textnormal{m}$ between the particle's
lower surface and the cathode's surface. The choice of $z_{d}^{\prime}=30\,\mu\textnormal{m}$
in Eq. (\ref{eq:IC-negative}) ensures such criteria. The parallel
plate electric field strength of $E_{p}=2\,\textnormal{MV}\cdot\textnormal{m}^{-1}$
is chosen. For all other parameter values, the same values from Eq.
(\ref{eq:parameter-LQ}) are used. That said, the result is plotted
in Fig. \ref{fig:11}, where it shows the oscillation frequency and
the oscillation amplitude identical to the positively charged particle
case corresponding to $E_{p}=2\,\textnormal{MV}\cdot\textnormal{m}^{-1}$
in Fig. \ref{fig:9}. This time, however, the charged-particle oscillates
near the cathode instead of the anode, as it is negatively charged. 

\subsubsection{Potential energy}

The potential energy for a negatively charged particle subjected to
the parallel plate electric field, $\mathbf{E}_{p},$ is obtained
from Eq. (\ref{eq:path-integral}) utilizing the line integral path
illustrated in Fig. \ref{fig:7}(b), 
\[
U\left(s\right)=-\int_{s_{\textnormal{ref}}}^{s}\mathbf{F}_{T}\cdot\mathbf{e}_{z}ds
\]
 or 
\begin{equation}
U=-\int_{r_{2}}^{r_{1}}\mathbf{F}_{T}\cdot\mathbf{e}_{z}ds,\label{eq:U-1}
\end{equation}
 where, for convenience, $s_{\textnormal{ref}}$ and $s$ have been
replaced by $r_{1}$ and $r_{2}$ illustrated in Fig. \ref{fig:7}(b),
respectively. Equivalently, Eq. (\ref{eq:U-1}) can be rewritten as
\begin{equation}
U=\int_{r_{1}}^{r_{2}}\mathbf{F}_{T}\cdot\mathbf{e}_{z}ds,\label{eq:U-2}
\end{equation}
 where the upper and the lower integral limits have been reversed.
Multiplication of the both sides of Eq. (\ref{eq:U-2}) by a negative
one yields 
\begin{equation}
-U=-\int_{r_{1}}^{r_{2}}\mathbf{F}_{T}\cdot\mathbf{e}_{z}ds.\label{eq:U-3}
\end{equation}
 The right side of Eq. (\ref{eq:U-3}) can be obtained by making the
following replacements in Eq. (\ref{eq:U(s)-pre-1}), 
\[
s_{\textnormal{ref}}\longrightarrow r_{1}\;\textnormal{ and }\; s\longrightarrow r_{2},
\]
 which yields 
\begin{align}
 & \int_{r_{1}}^{r_{2}}\mathbf{F}_{T}\cdot\mathbf{e}_{z}ds\nonumber \\
 & =-\frac{\pi\epsilon_{0}\kappa_{3}\nu}{4}\left\{ \frac{\nu}{r_{2}}+\frac{\nu}{h-r_{2}}+\frac{\left[\gamma\left(b^{3}-a^{3}\right)-b^{3}\right]E_{p}}{2r_{2}^{2}}\right.\nonumber \\
 & \left.-\frac{\left[\gamma\left(b^{3}-a^{3}\right)-b^{3}\right]E_{p}}{2\left(h-r_{2}\right)^{2}}+8E_{p}r_{2}\right\} -mgr_{2}\nonumber \\
 & +\frac{\pi\epsilon_{0}\kappa_{3}\nu}{4}\left\{ \frac{\nu}{r_{1}}+\frac{\nu}{h-r_{1}}+\frac{\left[\gamma\left(b^{3}-a^{3}\right)-b^{3}\right]E_{p}}{2r_{1}^{2}}\right.\nonumber \\
 & \left.-\frac{\left[\gamma\left(b^{3}-a^{3}\right)-b^{3}\right]E_{p}}{2\left(h-r_{1}\right)^{2}}+8E_{p}r_{1}\right\} +mgr_{1}.\label{eq:r1-r2-integral}
\end{align}
 Insertion of Eq. (\ref{eq:r1-r2-integral}) into Eq. (\ref{eq:U-2})
yields 
\begin{align}
U & =-\frac{\pi\epsilon_{0}\kappa_{3}\nu}{4}\left\{ \frac{\nu}{r_{2}}+\frac{\nu}{h-r_{2}}+\frac{\left[\gamma\left(b^{3}-a^{3}\right)-b^{3}\right]E_{p}}{2r_{2}^{2}}\right.\nonumber \\
 & \left.-\frac{\left[\gamma\left(b^{3}-a^{3}\right)-b^{3}\right]E_{p}}{2\left(h-r_{2}\right)^{2}}+8E_{p}r_{2}\right\} -mgr_{2}\nonumber \\
 & +\frac{\pi\epsilon_{0}\kappa_{3}\nu}{4}\left\{ \frac{\nu}{r_{1}}+\frac{\nu}{h-r_{1}}+\frac{\left[\gamma\left(b^{3}-a^{3}\right)-b^{3}\right]E_{p}}{2r_{1}^{2}}\right.\nonumber \\
 & \left.-\frac{\left[\gamma\left(b^{3}-a^{3}\right)-b^{3}\right]E_{p}}{2\left(h-r_{1}\right)^{2}}+8E_{p}r_{1}\right\} +mgr_{1}.\label{eq:U(r1,r2)}
\end{align}
 For a negatively charged particle, $r_{1}=s$ and $r_{2}=s_{\textnormal{ref}};$
and, thus, Eq. (\ref{eq:U(r1,r2)}) becomes 
\begin{align}
U\left(s\right) & =-\frac{\pi\epsilon_{0}\kappa_{3}\nu}{4}\left\{ \frac{\nu}{s_{\textnormal{ref}}}+\frac{\nu}{h-s_{\textnormal{ref}}}+\frac{\left[\gamma\left(b^{3}-a^{3}\right)-b^{3}\right]E_{p}}{2s_{\textnormal{ref}}^{2}}\right.\nonumber \\
 & \left.-\frac{\left[\gamma\left(b^{3}-a^{3}\right)-b^{3}\right]E_{p}}{2\left(h-s_{\textnormal{ref}}\right)^{2}}+8E_{p}s_{\textnormal{ref}}\right\} -mgs_{\textnormal{ref}}\nonumber \\
 & +\frac{\pi\epsilon_{0}\kappa_{3}\nu}{4}\left\{ \frac{\nu}{s}+\frac{\nu}{h-s}+\frac{\left[\gamma\left(b^{3}-a^{3}\right)-b^{3}\right]E_{p}}{2s^{2}}\right.\nonumber \\
 & \left.-\frac{\left[\gamma\left(b^{3}-a^{3}\right)-b^{3}\right]E_{p}}{2\left(h-s\right)^{2}}+8E_{p}s\right\} +mgs.\label{eq:NP-U(s)}
\end{align}
 In terms of the $z_{d}$ parameter, i.e., $s=z_{d}+b,$ Eq. (\ref{eq:NP-U(s)})
becomes 
\begin{align*}
U\left(z_{d}\right) & =-\frac{\pi\epsilon_{0}\kappa_{3}\nu}{4}\left\{ \frac{\nu}{s_{\textnormal{ref}}}+\frac{\nu}{h-s_{\textnormal{ref}}}+\frac{\left[\gamma\left(b^{3}-a^{3}\right)-b^{3}\right]E_{p}}{2s_{\textnormal{ref}}^{2}}\right.\\
 & \left.-\frac{\left[\gamma\left(b^{3}-a^{3}\right)-b^{3}\right]E_{p}}{2\left(h-s_{\textnormal{ref}}\right)^{2}}+8E_{p}s_{\textnormal{ref}}\right\} -mgs_{\textnormal{ref}}\\
 & +\frac{\pi\epsilon_{0}\kappa_{3}\nu}{4}\left\{ \frac{\nu}{z_{d}+b}+\frac{\nu}{h-z_{d}-b}+\frac{\left[\gamma\left(b^{3}-a^{3}\right)-b^{3}\right]E_{p}}{2\left(z_{d}+b\right)^{2}}\right.\\
 & \left.-\frac{\left[\gamma\left(b^{3}-a^{3}\right)-b^{3}\right]E_{p}}{2\left(h-z_{d}-b\right)^{2}}+8E_{p}\left(z_{d}+b\right)\right\} \\
 & +mg\left(z_{d}+b\right).
\end{align*}
 As with the case of the positive particle, I shall set $s_{\textnormal{ref}}$
at the midway between the parallel plates, i.e., Eq. (\ref{eq:sref-pos}),
\[
s_{\textnormal{ref}}=\frac{h}{2};
\]
 and, this yields 
\begin{align}
U\left(z_{d}\right) & =\frac{\pi\epsilon_{0}\kappa_{3}\nu}{4}\left\{ \frac{\nu}{z_{d}+b}+\frac{\nu}{h-z_{d}-b}\right.\nonumber \\
 & +\frac{\left[\gamma\left(b^{3}-a^{3}\right)-b^{3}\right]E_{p}}{2\left(z_{d}+b\right)^{2}}-\frac{\left[\gamma\left(b^{3}-a^{3}\right)-b^{3}\right]E_{p}}{2\left(h-z_{d}-b\right)^{2}}\nonumber \\
 & \left.+8E_{p}\left(z_{d}+b\right)\vphantom{\frac{\nu}{z_{d}+b}}\right\} +mg\left(z_{d}+b\right)\nonumber \\
 & -\frac{\pi\epsilon_{0}\kappa_{3}\nu}{h}\left(\nu+E_{p}h^{2}\right)-\frac{1}{2}mgh,\label{eq:NP-U(zd)-FINAL}
\end{align}
 where the resulting expression is, form wise, identical to the one
in Eq. (\ref{eq:U(zd)-FINAL}), i.e., the result corresponding to
the case of positive particle. 

\begin{figure}[h]
\begin{centering}
\includegraphics[width=0.85\columnwidth]{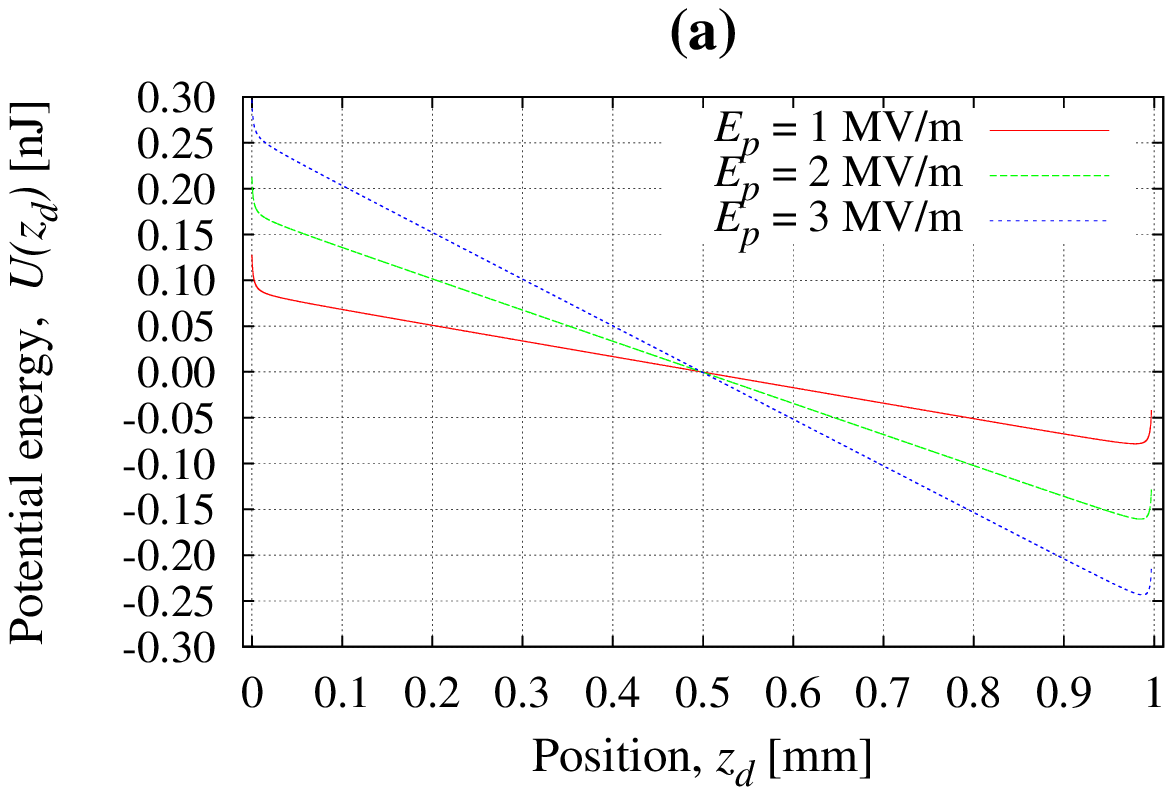}
\par\end{centering}

\begin{centering}
\includegraphics[width=0.85\columnwidth]{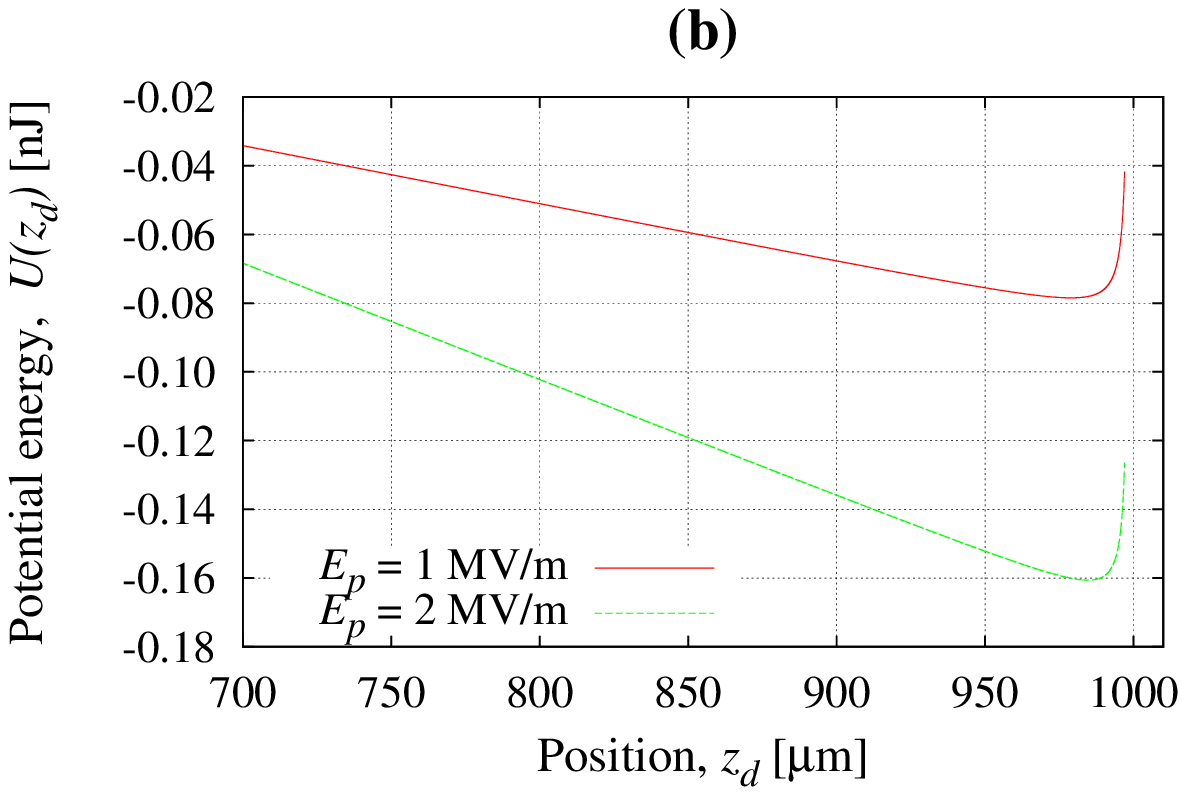}
\par\end{centering}

\caption{(Color online) (a) Plot of the potential energy function, $U\left(z_{d}\right)$
of Eq. (\ref{eq:NP-U(zd)-FINAL}), for $E_{p}=1\,\textnormal{MV}\cdot\textnormal{m}^{-1},$
$E_{p}=2\,\textnormal{MV}\cdot\textnormal{m}^{-1},$ and $E_{p}=3\,\textnormal{MV}\cdot\textnormal{m}^{-1}.$
The particle has a charge density of $\sigma_{1}=-0.012\,\textnormal{C}\cdot\textnormal{m}^{-2}$
and all other parameter values are as defined in Eq. (\ref{eq:parameter-LQ}).
(b) Enlarged plot of $U\left(z_{d}\right)$ for domain $h/2<z_{d}<h,$
where $h=1\,\textnormal{mm}.$ In the plot, the anode is located at
$z_{d}=0\,\mu\textnormal{m}.$ \label{fig:12}}
\end{figure}

Equation (\ref{eq:NP-U(zd)-FINAL}) is plotted for $\sigma_{1}=-0.012\,\textnormal{C}\cdot\textnormal{m}^{-2}$
with all other parameter values same as defined in Eq. (\ref{eq:parameter-LQ}),
\[
\left\{ \begin{array}{c}
\kappa_{2}=6,\quad\kappa_{3}=1,\\
a=1.5\,\mu\textnormal{m},\quad h=1\,\textnormal{mm},\\
b-a=4\,\textnormal{nm},\\
V_{L}=0\,\textnormal{V},\\
\sigma_{2}=0\,\textnormal{C}\cdot\textnormal{m}^{-2}\textnormal{ (i.e., insulator)},\\
\rho_{m,1}=2700\,\textnormal{kg}\cdot\textnormal{m}^{-3},\\
\rho_{m,2}=3800\,\textnormal{kg}\cdot\textnormal{m}^{-3}.
\end{array}\right.
\]
 For the plot, the anode voltages of $V_{T}=1\,\textnormal{kV},$
$V_{T}=2\,\textnormal{kV},$ and $V_{T}=3\,\textnormal{kV}$ are considered
for comparison. For $V_{L}=0\,\textnormal{V}$ and $h=1\,\textnormal{mm},$
these anode voltages correspond to $E_{p}=1\,\textnormal{MV}\cdot\textnormal{m}^{-1},$
$E_{p}=2\,\textnormal{MV}\cdot\textnormal{m}^{-1},$ and $E_{p}=3\,\textnormal{MV}\cdot\textnormal{m}^{-1},$
respectively. The results are shown in Fig. \ref{fig:12}. 

From the physical arguments based on the constituent forces, the oscillatory
solutions for a negatively charged particle occur only in the domain
$z_{d}>h/2$ in Figs. \ref{fig:12}(a) and Fig. \ref{fig:12}(b).
The results show the potential well minima occurring in vicinity of
the cathode side of the electrodes. Because the physical charged-particle
cannot penetrate into the surface of the cathode, the parameter $z_{d}$
in Fig. \ref{fig:12}(b) is bounded by $0\leq z_{d}\leq h-2b.$ The
case of $z_{d}=h-2b$ corresponds to a situation in which the charged-particle
is in physical contact with the cathode's surface. There, the height
of the potential well is finite and that criteria limits the width
of the potential well, wherein the negatively charged particle can
have oscillatory solutions. For instance, in the case of $E_{p}=1\,\textnormal{MV}\cdot\textnormal{m}^{-1}$
in Fig. \ref{fig:12}(b), the width of the potential well is approximately
$\sim250\,\mu\textnormal{m};$ and, the oscillatory solutions exist
approximately for $745\,\mu\textnormal{m}\lesssim z_{d}<996.99\,\mu\textnormal{m}.$
Similarly, for the case of $E_{p}=2\,\textnormal{MV}\cdot\textnormal{m}^{-1}$
in Fig. \ref{fig:12}(b), the width of the potential well is approximately
$\sim125\,\mu\textnormal{m};$ and, the negatively charged particle
is expected to have oscillatory solutions in a domain $\sim872\,\mu\textnormal{m}\lesssim z_{d}<996.99\,\mu\textnormal{m}.$
For the negatively charged particle, the potential well, wherein the
oscillatory solutions exist, gets formed in vicinity of the cathode.
The width of such potential well decreases with increased $E_{p}.$
Physically, such property corresponds to the narrowing of the negative
glow region with increased $E_{p}.$

\subsection{Dipole radiation}

\subsubsection{Nonrelativistic regime}

Oscillating charged-particle radiates electromagnetic energy. The
power of such radiation, in the nonrelativistic limit, is given by
Larmor radiation formula, 
\begin{align}
P_{rad}\left(t\right) & =\frac{Q_{T}^{2}\ddot{z}_{d}^{2}}{6\pi\epsilon_{0}c^{3}},\label{eq:larmor-formula}
\end{align}
 where $\ddot{z}_{d}$ is the nonrelativistic charged-particle acceleration
defined in Eq. (\ref{eq:ODE-newton}), and $Q_{T}$ is the effective
charge carried by the particle,\cite{Cho} 
\begin{equation}
Q_{T}=4\pi\epsilon_{0}\kappa_{3}\nu.\label{eq:QT-in-terms-of-nu}
\end{equation}
 Insertion of Eqs. (\ref{eq:ODE-newton}) and (\ref{eq:QT-in-terms-of-nu}),
respectively for $\ddot{z}_{d}$ and $Q_{T},$ into Eq. (\ref{eq:larmor-formula})
yields 
\begin{align}
P_{rad}\left(t\right) & =\frac{8\pi\epsilon_{0}\kappa_{3}^{2}\nu^{2}}{3c^{3}}\left(\frac{\pi\epsilon_{0}\kappa_{3}\nu}{4m}\left\{ \frac{\nu}{\left(z_{d}+b\right)^{2}}\right.\right.\nonumber \\
 & -\frac{\nu}{\left(h-z_{d}-b\right)^{2}}+\frac{\left[\gamma\left(b^{3}-a^{3}\right)-b^{3}\right]E_{p}}{\left(z_{d}+b\right)^{3}}\nonumber \\
 & \left.\left.+\frac{\left[\gamma\left(b^{3}-a^{3}\right)-b^{3}\right]E_{p}}{\left(h-z_{d}-b\right)^{3}}-8E_{p}\right\} -g\right)^{2},\label{eq:larmor-power}
\end{align}
 where $z_{d}$ is the solution to the Eq. (\ref{eq:ODE-newton}). 

\begin{figure*}[t]
\begin{centering}
\includegraphics[width=0.9\columnwidth]{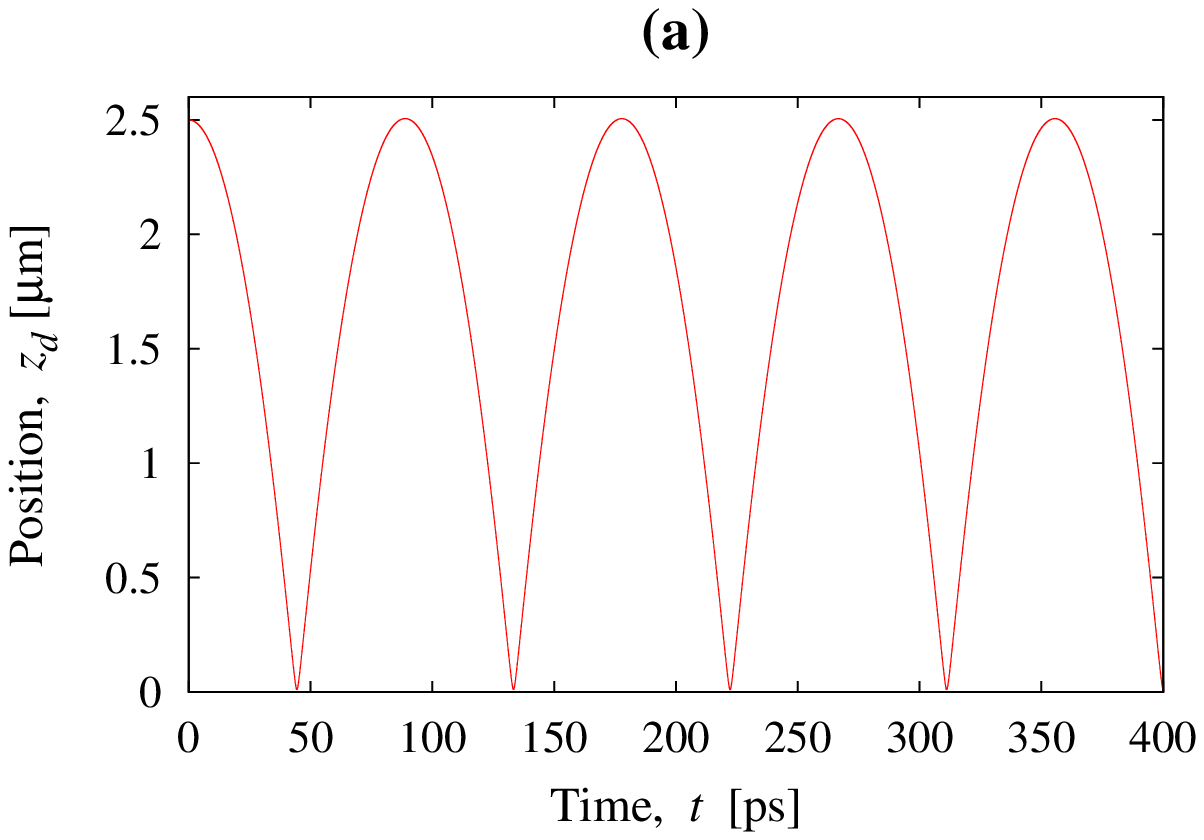}\includegraphics[width=0.9\columnwidth]{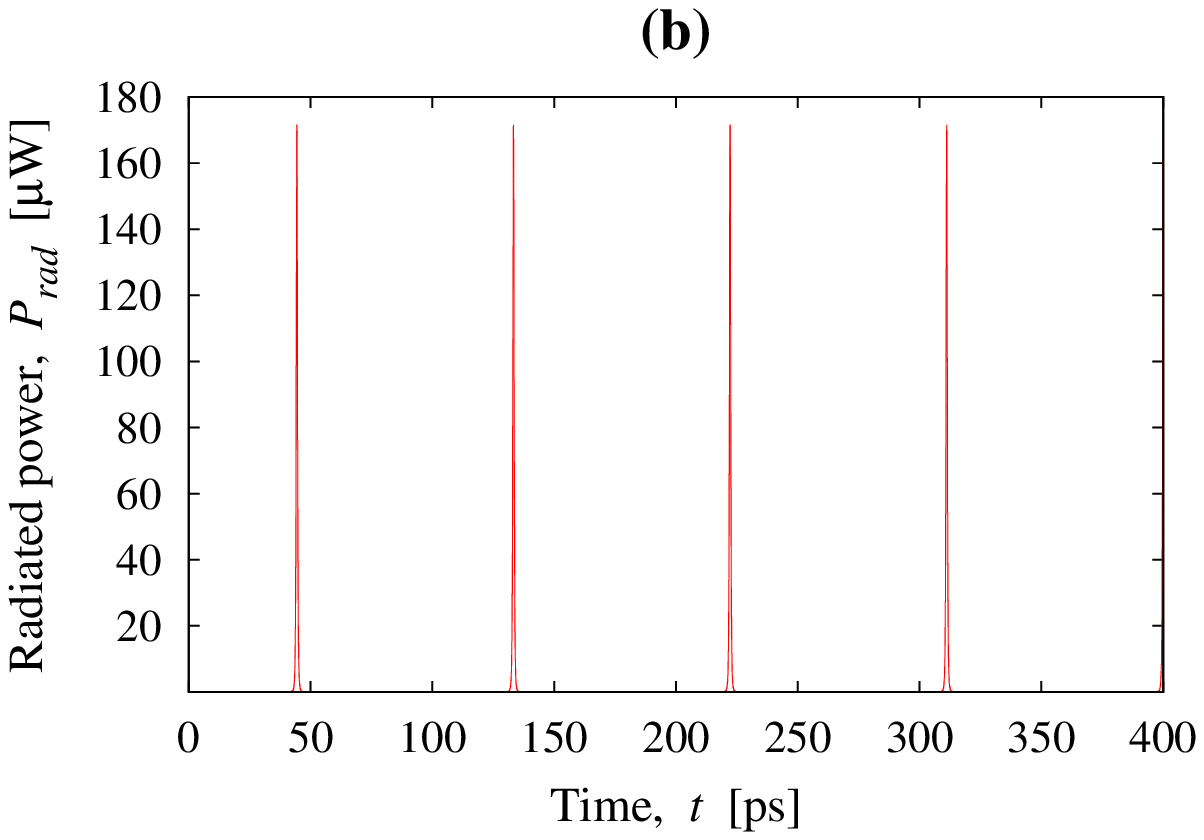}
\par\end{centering}

\begin{centering}
\includegraphics[width=0.9\columnwidth]{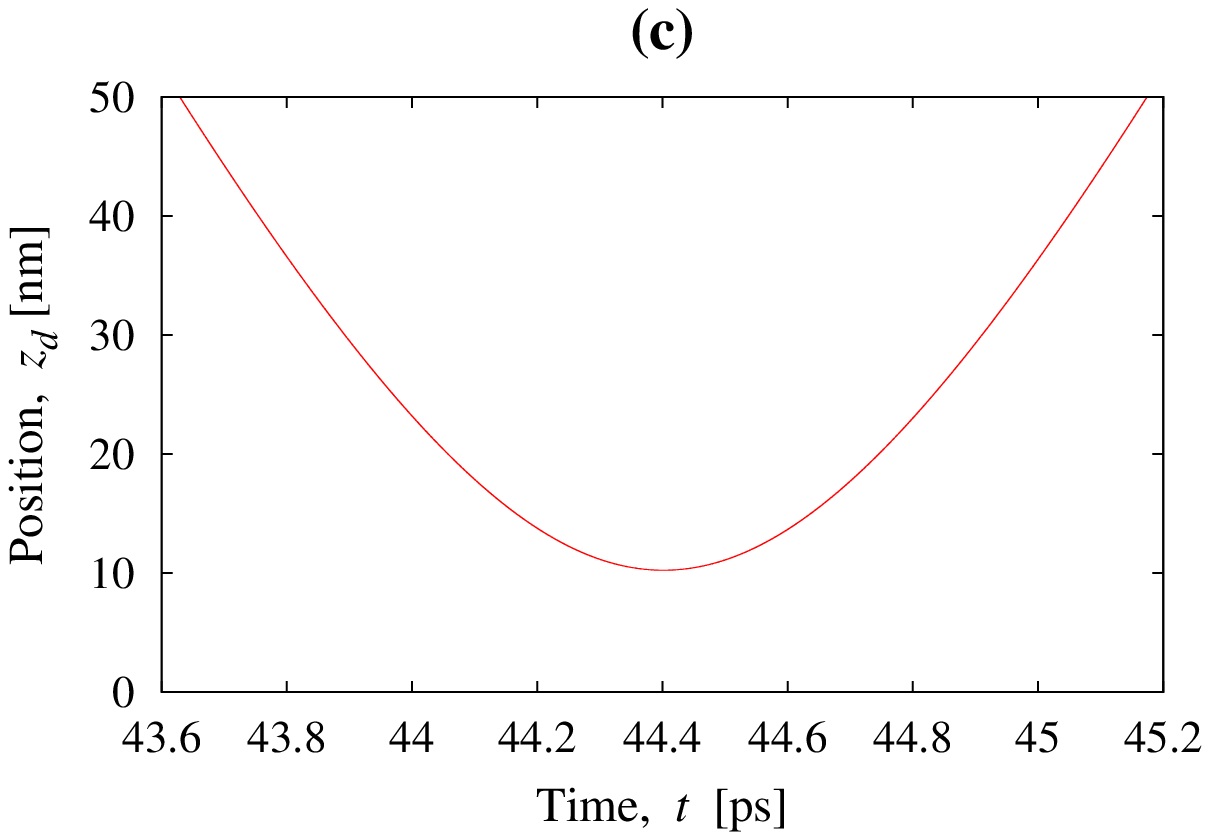}\includegraphics[width=0.9\columnwidth]{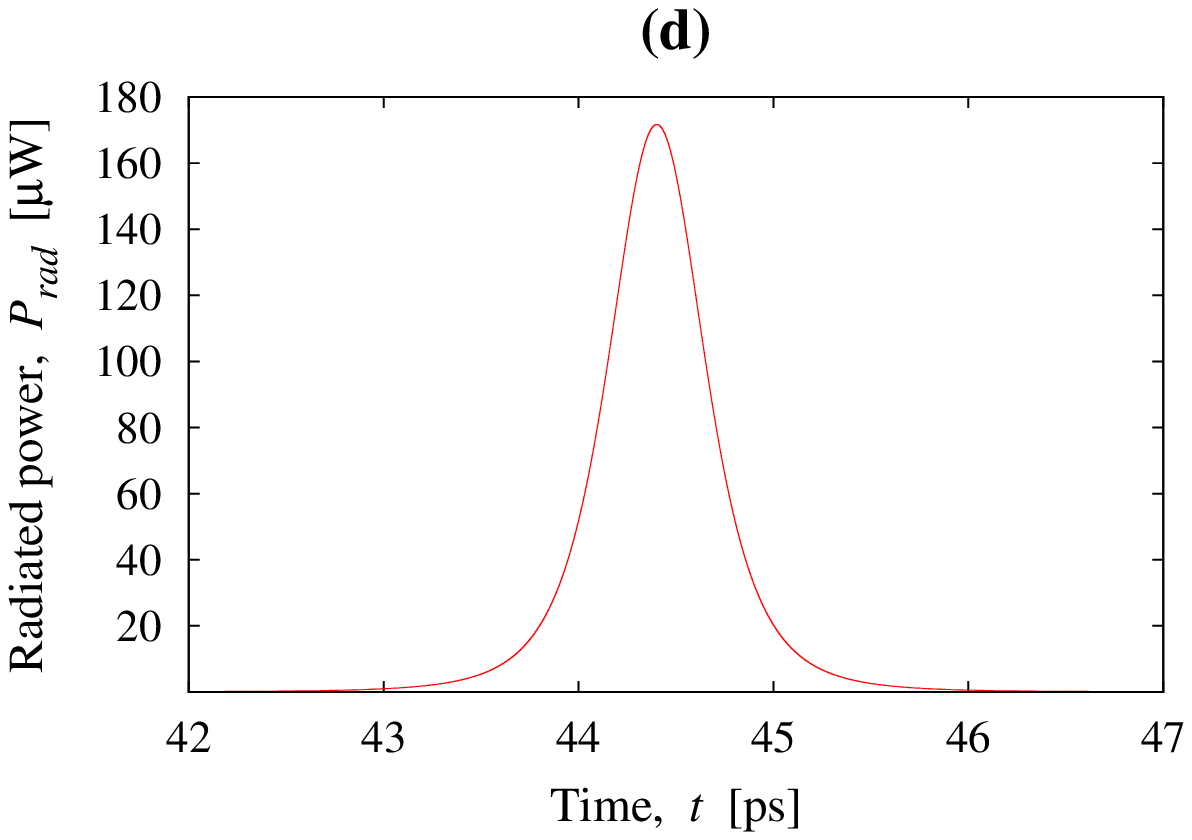}
\par\end{centering}

\caption{(Color online) (a) Plot of $z_{d}\left(t\right),$ Eq. (\ref{eq:ODE-newton}),
using the parameter values defined in Eq. (\ref{eq:parameter-HQ})
and the initial conditions defined in Eq. (\ref{eq:IC-positive-HQ}).
(b) The corresponding Larmor radiation power computed from Eq. (\ref{eq:larmor-power}).
(c) The first sharp turning point in (a) has been zoomed for a detailed
view. (d) The first pulse in (b) has been zoomed for a detailed view.
In the plot, the anode plate is located at $z_{d}=0\,\mu\textnormal{m}.$
\label{fig:13}}
\end{figure*}

I shall use the following parameter values to evaluate $P_{rad}\left(t\right):$
\begin{equation}
\left\{ \begin{array}{c}
\kappa_{2}=6,\quad\kappa_{3}=1,\\
a=25\,\textnormal{nm},\quad h=10\,\mu\textnormal{m},\\
b-a=2\,\textnormal{nm},\\
V_{T}=16\,\textnormal{kV},\quad V_{L}=0\,\textnormal{V},\\
\sigma_{1}=100\,\textnormal{C}\cdot\textnormal{m}^{-2},\\
\sigma_{2}=0\,\textnormal{C}\cdot\textnormal{m}^{-2}\textnormal{ (i.e., insulator)},\\
\rho_{m,1}=2700\,\textnormal{kg}\cdot\textnormal{m}^{-3},\\
\rho_{m,2}=3800\,\textnormal{kg}\cdot\textnormal{m}^{-3}.
\end{array}\right.\label{eq:parameter-HQ}
\end{equation}
 The corresponding charged-particle motion is obtained by solving
Eq. (\ref{eq:ODE-newton}) via Runge-Kutta method. For the initial
conditions, I shall choose 
\begin{equation}
z_{d}\left(0\right)=1\,\mu\textnormal{m}\quad\textnormal{and}\quad\dot{z}_{d}\left(0\right)=0.\label{eq:IC-positive-HQ}
\end{equation}
 Illustrated in Figs. \ref{fig:13}(a) and \ref{fig:13}(b) are the
results of $z_{d}\left(t\right)$ and $P_{rad}\left(t\right),$ respectively.
The first sharp turning point in the plot of Fig. \ref{fig:13}(a)
has been enlarged and is shown in Fig. \ref{fig:13}(c), where it
shows the particle rebounding approximately at a distance of $10\,\textnormal{nm}$
from the anode plate electrode's surface. The first pulse of the emitted
Larmor radiation power has been enlarged and is shown in Fig. \ref{fig:13}(d). 

Is $\sigma_{1}=100\,\textnormal{C}\cdot\textnormal{m}^{-2}$ experimentally
an attainable surface charge density? The answer to this question
is yes. In fact, for systems wherein nanoparticles are deliberately
ionized in controlled manner, the surface charge density of $\sigma_{1}=100\,\textnormal{C}\cdot\textnormal{m}^{-2}$
is more reasonable than the one used in Eq.  (\ref{eq:parameter-LQ}),
i.e., $\sigma_{1}=0.012\,\textnormal{C}\cdot\textnormal{m}^{-2}.$
Physically, $\sigma_{1}=100\,\textnormal{C}\cdot\textnormal{m}^{-2}$
corresponds to a case wherein each aluminum atoms in the volume of
radius $r=a$ contributing approximately one electron in the ionization
process. This can be illustrated as follow. The mass of spherical
aluminum core is 
\[
m_{c}=\frac{4}{3}\pi a^{3}\rho_{m,1};
\]
 and, the mass of a single aluminum atom is given by 
\[
m_{al}=\frac{m_{w}}{N_{A}},
\]
 where $m_{w}$ is the molar atomic weight and $N_{A}\approx6.022\times10^{23}\,\textnormal{mol}^{-1}$
is the Avogadro constant. The total number of aluminum atoms inside
the volume of radius $r=a$ can be calculated as 
\[
N_{al}=\frac{m_{c}}{m_{al}}=\frac{4N_{A}\pi a^{3}\rho_{m,1}}{3m_{w}}.
\]
 The aluminum core of radius $r=a$ carries a surface charge of 
\[
Q_{sc}=4\pi a^{2}\sigma_{1}.
\]
 For a macroscopic particle, $Q_{sc}$ would be solely contributed
from the atoms near the surface. However, for nanoparticles, the idea
of ``surface charge'' becomes vague because it's not just those
atoms near the surface, but the atoms in the entire volume of nanoparticle
that contribute to $Q_{sc}.$ In that sense, $Q_{sc}$ should be more
appropriately coined as the ``total charge'' carried by the nanoparticle,
\textit{albeit} it is still defined in terms of the surface charge
formula, $Q_{sc}=4\pi a^{2}\sigma_{1}.$ That said, how many electrons
must be removed from each aluminum atoms in order for the core to
have net positive charge in the amount of $Q_{sc}=4\pi a^{2}\sigma_{1}$?
The answer to this question is 
\[
N_{e}=\frac{Q_{sc}}{q_{e}N_{al}}=\frac{3\pi\sigma_{1}m_{w}}{q_{e}N_{A}\pi a\rho_{m,1}},
\]
 where $q_{e}\approx1.602\times10^{-19}\,\textnormal{C}$ is the fundamental
charge magnitude. For an aluminum atom, $m_{w}\approx26.98\,\textnormal{g}$
and the number of electrons to be removed per aluminum atom is 

\[
N_{e}\approx1.24\;\longrightarrow\; N_{e}=1,
\]
 where the greatest integer value has been taken for $N_{e},$ as
there cannot be $1.24$ electrons, of course. This result implies
that, on the average, each aluminum atoms in the core loses one electron
during the ionization process in the case of $\sigma_{1}=100\,\textnormal{C}\cdot\textnormal{m}^{-2}$
and $a=25\,\textnormal{nm}.$ 

The anode voltage of $V_{T}=16\,\textnormal{kV}$ has been carefully
chosen such that electrical breakdown does not occur between the parallel
plates. Zouache and Lefort have demonstrated that by choosing a composite
material for electrodes, for instance, composite material of $60\textnormal{\%}$
silver and $40\textnormal{\%}$ nickel, the DC bias voltage across
the two electrodes can be as high as $3.85\,\textnormal{kV}$ at plate
gap of $1\,\mu\textnormal{m}$ in vacuum before electrical breakdown
takes place.\cite{IEEE-vacuum} In terms of electric field strength,
this corresponds to $E_{p}=3.85\,\textnormal{GV}\cdot\textnormal{m}^{-1}.$
At plate gap of $h=10\,\mu\textnormal{m}$ and the cathode grounded,
the anode voltage of $V_{T}=16\,\textnormal{kV}$ corresponds to $E_{p}=1.6\,\textnormal{GV}\cdot\textnormal{m}^{-1},$
which is much less than $E_{p}=3.85\,\textnormal{GV}\cdot\textnormal{m}^{-1}.$

\subsubsection{Relativistic regime}

The Larmor radiation formula, Eq. (\ref{eq:larmor-formula}), is only
valid for particle speeds that are small relative to the speed of
light. In the relativistic generalization, the total power radiated
by oscillating charged-particle is given by the Liénard radiation
formula,\cite{Cho} 
\begin{align}
P_{rad}\left(t\right) & =\frac{8\pi\epsilon_{0}\kappa_{3}^{2}\nu^{2}}{3c^{3}}\left[1-\left(\frac{\dot{z}_{d}}{c}\right)^{2}\right]^{-3}\ddot{z}_{d}^{2},\label{eq:lienard-formula}
\end{align}
 where $\ddot{z}_{d}$ is the particle's acceleration associated with
the relativistic force, Eq. (\ref{eq:einstein}). With the explicit
expression for $\ddot{z}_{d}$ inserted from Eq. (\ref{eq:ODE-einstein-fin}),
the Liénard radiation formula of Eq. (\ref{eq:lienard-formula}) becomes 

\begin{align}
P_{rad}\left(t\right) & =\frac{8\pi\epsilon_{0}\kappa_{3}^{2}\nu^{2}}{3c^{3}}\left(\frac{\pi\epsilon_{0}\kappa_{3}\nu}{4m}\left\{ \frac{\nu}{\left(z_{d}+b\right)^{2}}\right.\right.\nonumber \\
 & -\frac{\nu}{\left(h-z_{d}-b\right)^{2}}+\frac{\left[\gamma\left(b^{3}-a^{3}\right)-b^{3}\right]E_{p}}{\left(z_{d}+b\right)^{3}}\nonumber \\
 & \left.\left.+\frac{\left[\gamma\left(b^{3}-a^{3}\right)-b^{3}\right]E_{p}}{\left(h-z_{d}-b\right)^{3}}-8E_{p}\right\} -g\right)^{2},\label{eq:lienard-power}
\end{align}
 which expression is identical to Eq. (\ref{eq:larmor-power}) with
an exception that $z_{d}$ is now obtained from the relativistic dynamics,
i.e., Eqs. (\ref{eq:einstein}) or (\ref{eq:ODE-einstein-fin}).

\section{\noindent Mechanism for self-sustained oscillations in the positive
glow corona}

\begin{figure}[h]
\begin{centering}
\includegraphics[width=1\columnwidth]{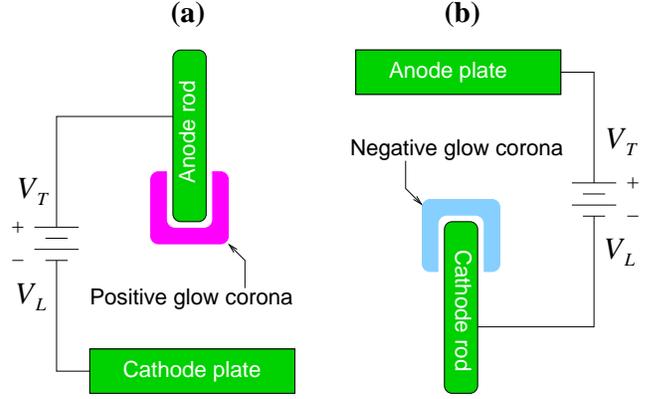}
\par\end{centering}

\caption{(Color online) (a) Schematic illustration of positive glow corona
apparatus. In the positive glow corona, a glowing light is observed
in vicinity of the anode. (b) Schematic illustration of a negative
glow corona. In the negative glow corona, a glowing light is observed
in vicinity of the cathode. \label{fig:14}}
\end{figure}

The typical configurations for the electrodes in the glow corona apparatus
are illustrated in Fig. \ref{fig:14}. When the potential difference
between the rod shaped electrode and the plate electrode are sufficiently
large, but not large enough to cause an electric arc, a glow occurs
near the surface of the rod shaped electrode. For the case in which
the rod shaped electrode is an anode, the glow corona is referred
to as the positive glow corona whereas, if the rod shaped electrode
is the cathode, the glow corona is referred to as the negative glow
corona by convention. The schematics of the positive and the negative
glow corona apparatuses are illustrated in Figs. \ref{fig:14}(a)
and \ref{fig:14}(b), respectively. In this paper, only the positive
glow corona is discussed. 

The first reported account with the glow corona is by Michael Faraday
in 1838. However, only recently, people have begun to extensively
investigate the phenomenon due to its potential applications in various
scientific and engineering fields, such as semiconductor lithography,
materials processing, plasma lighting, and so on.\cite{XUV} Among
the interesting properties of the phenomenon of glow corona, the self-sustained
electrode current oscillations in the positive glow corona is the
most extensively investigated one. Such self-sustained oscillations
in the electrode current persists \textit{albeit} the system as a
whole is biased with a DC voltage across the electrodes. To this date,
the basic underlying mechanism behind such self-sustained current
oscillations remains unclear.\cite{corona-discharge-1,corona-discharge-2,corona-discharge-3,corona-discharge-4,corona-discharge-5} 

\begin{figure}[h]
\begin{centering}
\includegraphics[width=0.9\columnwidth]{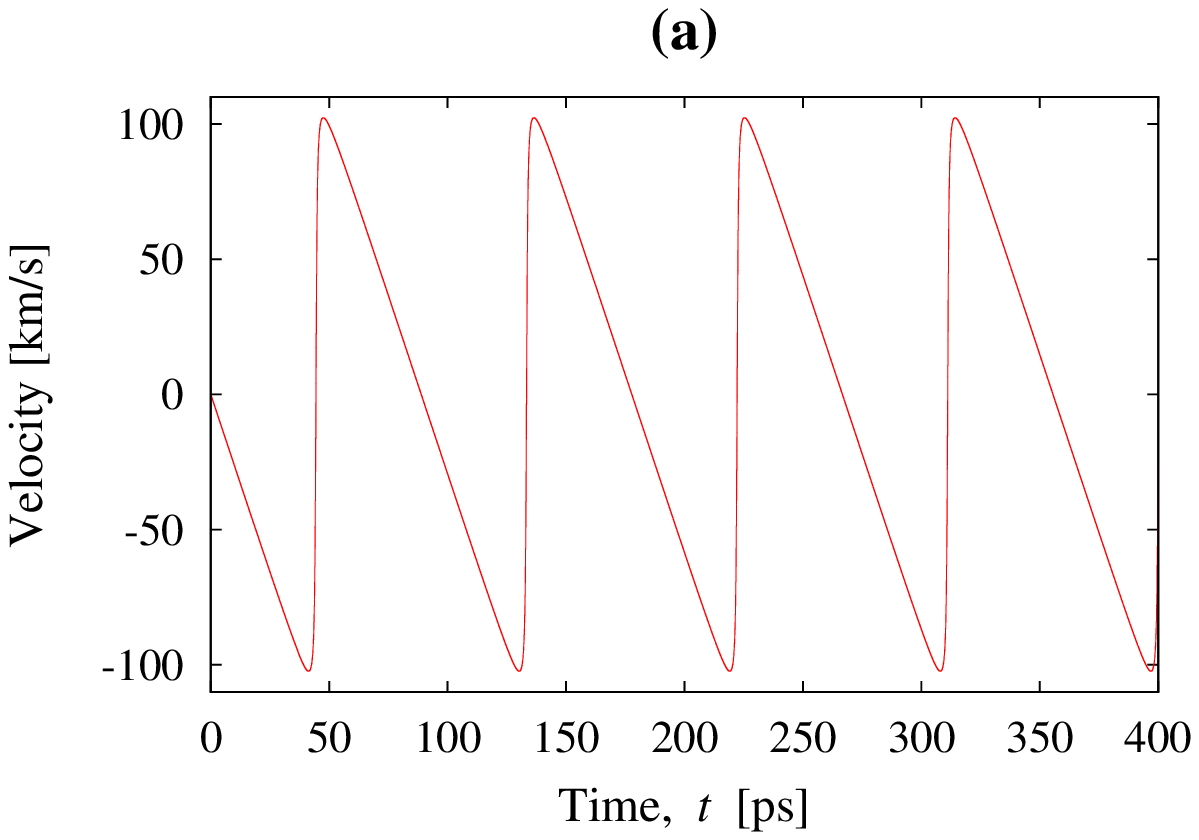}
\par\end{centering}

\begin{centering}
\includegraphics[width=0.9\columnwidth]{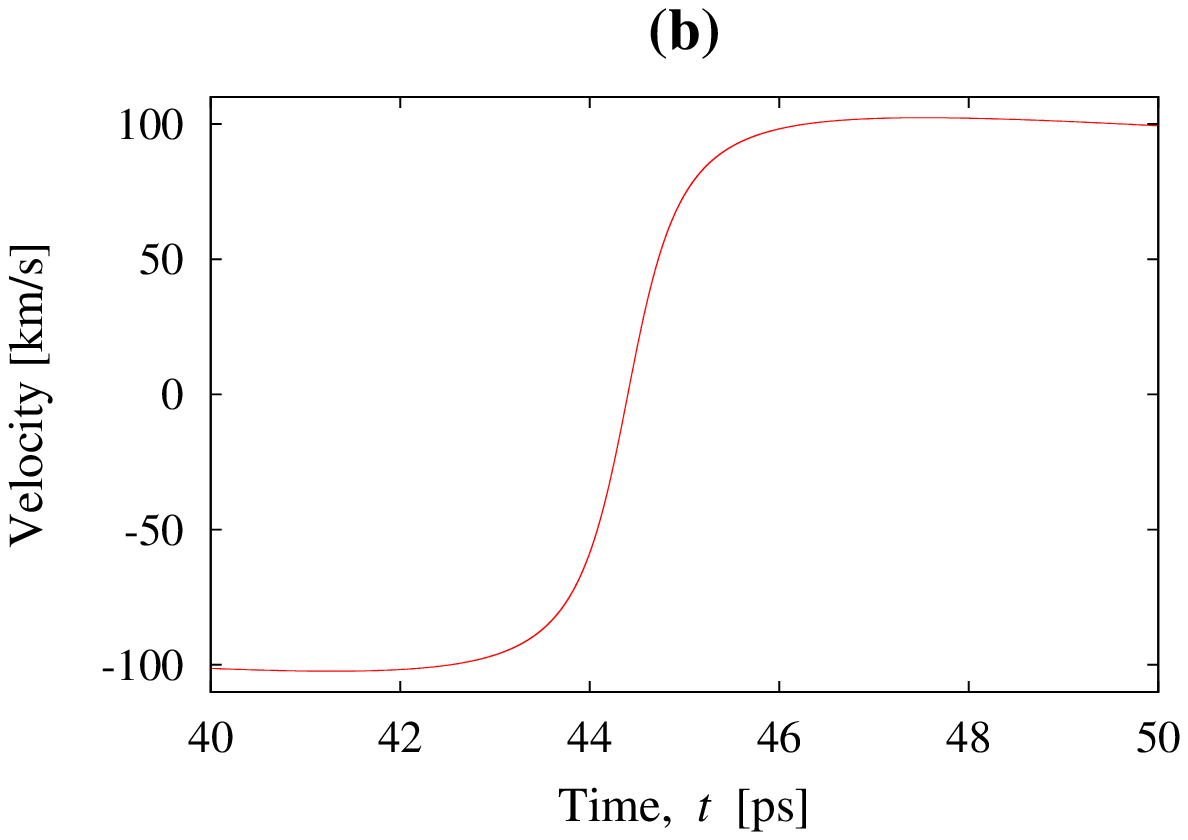}
\par\end{centering}

\caption{(Color online) (a) The plot of velocity, $\dot{\mathbf{z}}_{d}\left(t\right),$
corresponding to the charged-particle in Fig. \ref{fig:13}. (b) The
first abrupt rise in the velocity has been zoomed for a detailed view.
\label{fig:15}}
\end{figure}

Morrow did a theoretical work in an attempt to explain the underlying
physics behind the self-sustained current oscillations in the positive
glow corona.\cite{corona-discharge-3} Qualitatively, his predictions
are consistent with the various experimental observations by others.
Despite the fact that plasmas are ionized gases,\cite{Bogaerts-DC-glow-discharge,Gyergyek}
which may contain particles of all charged species (positive, negative,
or neutral) of various sizes (atoms or nanoparticles), Morrow showed
that the self-sustained oscillations in the electrode current are
predominantly due to the mobility of the positive ions in the gas.
He has calculated that approximately $96\%$ of the variations in
the electrode current is due to the oscillations of the positive ions
in the plasma. The current in the electrode can thus be expressed
as 
\[
\mathbf{I}\left(t\right)\propto\sum_{n}Q_{T,n}\dot{\mathbf{z}}_{d,n}\left(t\right),
\]
 where $Q_{T,n}$ and $\dot{\mathbf{z}}_{d,n}\left(t\right)$ are
the effective charge and the velocity of the $n\textnormal{th}$ charged-particle,
respectively. In this paper, there is only one charged-particle; and,
the expression for the electrode current becomes 
\[
\mathbf{I}\left(t\right)\propto Q_{T}\dot{\mathbf{z}}_{d}\left(t\right),
\]
 where $Q_{T}$ is the effective charge carried by the charged-particle
and $\dot{\mathbf{z}}_{d}\left(t\right)$ is the charged-particle's
velocity. For the core-shell structured charged-particle considered
here, the expression for $Q_{T}$ is\cite{Cho} 
\[
Q_{T}=8\pi a\left(b-a\right)\sigma_{1}\frac{\kappa_{3}}{\kappa_{2}}+4\pi\left(a^{2}\sigma_{1}+b^{2}\sigma_{2}\right),
\]
 which is a constant. Neglecting the constant terms, the current in
the electrode becomes 
\begin{equation}
\mathbf{I}\left(t\right)\propto\dot{\mathbf{z}}_{d}\left(t\right).\label{eq:J-in-velocity}
\end{equation}
 The $\dot{\mathbf{z}}_{d}\left(t\right)$ has been computed for the
nonrelativistic charged-particle whose position versus time plot and
the associated Larmor radiation power are illustrated in Figs. \ref{fig:13}(a)
and \ref{fig:13}(b). The result is shown in Fig. \ref{fig:15}(a),
where the first abrupt rise in the velocity has been zoomed for a
detailed view in Fig. \ref{fig:15}(b). This result is compared with
waveforms of experimental discharge current measurement at the electrode
for a positive corona in nitrogen at 35 Torr by Akishev \textit{et
al}.\cite{corona-discharge-4} Similarly, the result is also compared
with the prediction by Morrow.\cite{corona-discharge-3} Remarkably,
the profile of Eq. (\ref{eq:J-in-velocity}), which is shown in Fig.
\ref{fig:15}(a), closely resembles both experimental and theoretical
results by Akishev \textit{et al}. and Morrow, respectively. For instance,
the current in Eq. (\ref{eq:J-in-velocity}) has a saw-tooth shaped
wave profile, which is qualitatively similar to the waveforms obtained
by Akishev \textit{et al}. and Morrow. Moreover, presented theory
predicts pulses of radiation output occurring precisely at the point
where the current rises abruptly. This can be checked from Figs. \ref{fig:13}(b)
and \ref{fig:15}(a), where it shows pulses of radiation power coinciding
with the abrupt rises in the charged-particle velocity versus time
plot. Such characteristic is consistent with results obtained by Akishev
\textit{et al}. and Morrow. This shows that positive ion oscillations
in the positive glow corona involve the kind of charged-particle oscillation
mechanism discussed in Fig. \ref{fig:3}. 

\begin{figure*}[t]
\begin{centering}
\includegraphics[width=1.4\columnwidth]{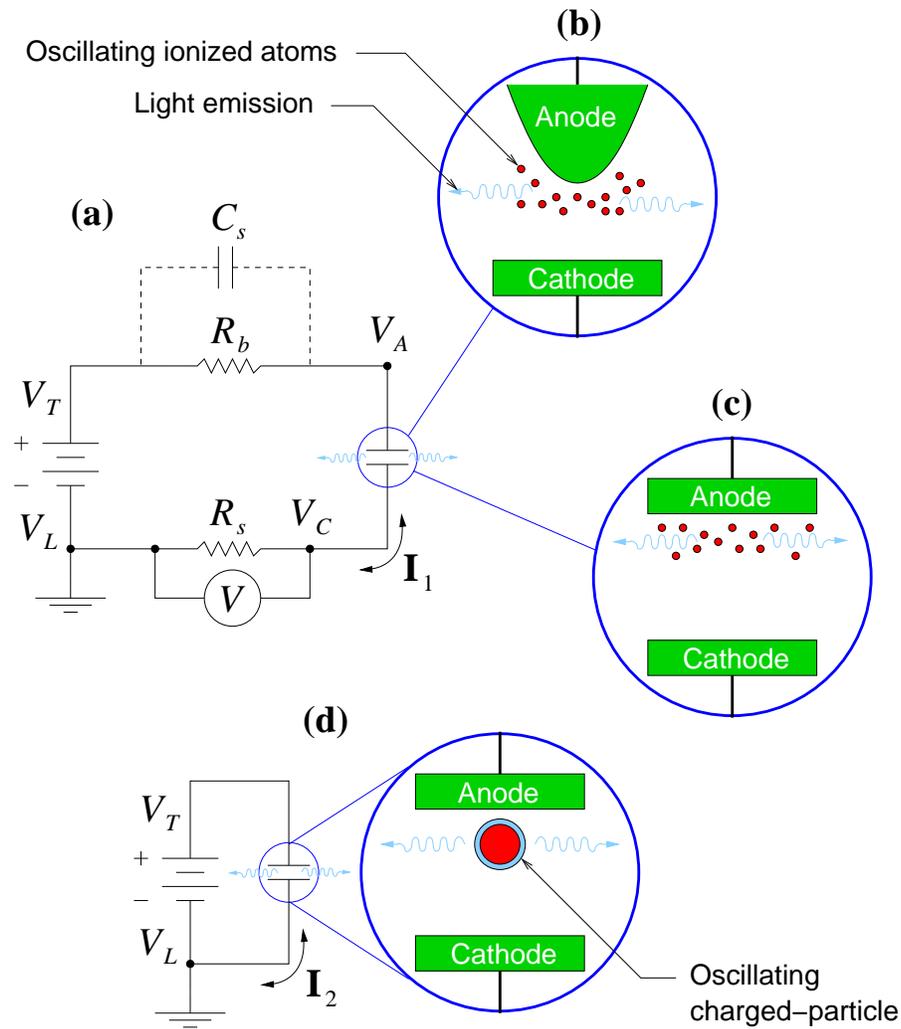}
\par\end{centering}

\caption{(Color online) (a) Typical experimental apparatus in the glow corona
(or glow discharge) experiment. (b) Electrodes in the positive glow
corona experiment. (c) Electrodes in the DC glow discharge experiment
using parallel plates. The ballast resistor $R_{b}$ and the shunt
resistor $R_{s}$ are shown. The voltmeter is placed across the shunt
resistor. The $C_{s}$ is a stray capacity of external circuit. (d)
The equivalent circuit diagram corresponding to the model in this
paper.\label{fig:16}}
\end{figure*}

The minor discrepancies in the electrode current waveforms between
the result of this work and the experimental measurements by others
can be attributed to the differences in the setup of the apparatus.\cite{corona-discharge-1,corona-discharge-2,corona-discharge-3,corona-discharge-4,corona-discharge-5}
Illustrated in Fig. \ref{fig:16}(a) is the equivalent circuit diagram
representation for a typical apparatus in the glow corona experiments.
The setup for a typical glow corona experiment involves the ballast
resistor $R_{b},$ a shunt resistor $R_{s},$ and a stray capacitance
$C_{s}$ from the external circuit. In the positive glow corona experiment,\cite{corona-discharge-4}
the typical geometry for the anode and cathode electrodes are as illustrated
in Fig. \ref{fig:16}(b) whereas, in a typical DC glow discharge experiments,
the parallel plate geometry, such as the one illustrated in Fig. \ref{fig:16}(c),
is the typical configuration for the electrodes.\cite{Kuschel-axial-light}
The presence of the ballast and the shunt resistors in the circuit
keeps the anode at voltage $V_{A}\left(t\right)$ and the cathode
at voltage $V_{C}\left(t\right).$ 

These configurations for the glow corona experiment, Figs. \ref{fig:16}(a)-\ref{fig:16}(c),
are compared with the model configuration considered in this work,
Fig. \ref{fig:4}. The equivalent circuit diagram for the model illustrated
in Fig. \ref{fig:4} is as shown in Fig. \ref{fig:16}(d). Unlike
the typical setup in the glow corona (or DC glow discharge) experiments,
the equivalent circuit diagram for the model considered here does
not contain the ballast and the shunt resistors in the circuit. Absence
of these resistors in the circuit keep the anode and the cathode voltages
fixed, respectively, at $V_{T}$ and $V_{L}$ in Fig. \ref{fig:4}.
Contrary to this, the anode and the cathode voltages in a typical
glow corona experiment are not fixed at some constant values due to
the presence of the ballast and the shunt resistors. For instance,
the voltages $V_{A}\left(t\right)$ and $V_{C}\left(t\right)$ in
Fig. \ref{fig:16}(a) are not constants, but vary in time due to the
dynamics of ionized atoms in the space between the anode and the cathode
electrodes. For this reason, the electrode voltage oscillation measurements
from an experiment, in which the setup is equivalent to the one illustrated
in Fig. \ref{fig:16}(a), cannot be used directly to test the theory
presented in this paper. However, the current oscillations in the
electrodes are present in all of the configurations in Fig. \ref{fig:16}.
For instance, an oscillating ionized particle between the anode and
the cathode gives rise to an electrode current, $\mathbf{I}_{1}\left(t\right)$
in Fig. \ref{fig:16}(a) and $\mathbf{I}_{2}\left(t\right)$ in Fig.
\ref{fig:16}(d), which oscillates in correlation to the motion of
oscillating ionized particle. Such electrode current oscillations
get induced in the circuit regardless of whether the ballast and the
shunt resistors are present in the circuit or not. The electrode current
oscillation measurements from Figs. \ref{fig:16}(a)-\ref{fig:16}(c),
therefore, can be used to test the theory presented here; and, this
is essentially what was assumed in Eq. (\ref{eq:J-in-velocity}). 

Since the geometry of the anode used in the experiment is different
from the simple plate geometry assumed in the model adopted in this
paper, the resulting waveforms of oscillating electrode currents from
this theory, $\mathbf{I}_{2}\left(t\right),$ and the experiment,
$\mathbf{I}_{1}\left(t\right),$ are not identical. Nevertheless,
qualitatively, both $\mathbf{I}_{1}\left(t\right)$ and $\mathbf{I}_{2}\left(t\right)$
show the same characteristic behavior. This must be so because the
basic mechanism behind the oscillations in $\mathbf{I}_{1}\left(t\right)$
and $\mathbf{I}_{2}\left(t\right)$ originates from the same physics.
One such characteristic behavior is the presence of radiation output
accompanying the abrupt rises in the $\mathbf{I}_{1}\left(t\right)$
and $\mathbf{I}_{2}\left(t\right),$ which was discussed previously
from Figs. \ref{fig:13}(b) and \ref{fig:15}(a). 

Besides the geometrical differences in the anode, the model treated
here has only a single charged-particle in the space between the anode
and the cathode whereas, in the glow corona experiments,\cite{corona-discharge-1,corona-discharge-2,corona-discharge-4}
the space between the electrodes is filled with an ionized gas, i.e.,
many ionized atoms. Despite these differences, the theory qualitatively
reproduces the self-sustained current oscillations in the electrode,
consistent with the results from the various glow corona experiments.
Such result suggests that the phenomenon of self-sustained electrode
current oscillations in the positive glow corona is a manifestation
of the charged-particle oscillation discussed in this paper. 

The radiation power in Fig. \ref{fig:13}(b) is emitted at frequency
of approximately $10\,\textnormal{GHz},$ which is not a visible light.
How can the pulses of light accompanying the saw-tooth shaped electrode
current oscillations be explained? To answer this, typical glow corona
experiment involves gases. There, individual atoms can be highly ionized
to oscillate at frequencies large enough to emit visible light. The
plasma as a whole, however, oscillates at much smaller frequencies
because its dynamics involves the collective motions of all constituent
ionized atoms. This qualitatively explains the pulses of light accompanying
the self-sustained electrode current oscillations, which oscillates
at much lower frequencies. 

As an extension of this theory, the self-sustained oscillations in
the negative glow corona, Fig. \ref{fig:14}b, can be qualitatively
explained from the negatively charged particles going through an oscillatory
motion in vicinity of the cathode, which is schematically illustrated
in Fig. \ref{fig:10}. The neon lamp used in the Pearson\textendash{}Anson
relaxation oscillator is the classic example of negative glow corona
at work. When the neon bulb is biased with a direct current (DC) voltage,
typically around $60\,\textnormal{V},$ a glow gets formed around
the cathode lead. No such glow occurs near the anode lead. It is unlikely
that neon atoms inside the lamp are positively charged. Quantum mechanical
calculations show that it takes minimum electric field strength of
approximately $4.8\,\textnormal{V}\cdot\textnormal{Å}^{-1}$ (or $48\,\textnormal{GV}\cdot\textnormal{m}^{-1}$)
to strip an electron from a neon at temperature of $273^{\text{\textdegree}}\textnormal{K}.$\cite{field-emission-neon}
In a typical neon bulbs used in the Pearson-Anson relaxation oscillators,
the anode and the cathode leads are separated by a gap of just few
millimeters. Assuming a gap of $1\,\textnormal{mm}$ between the electrodes,
and a DC bias voltage of $60\,\textnormal{V},$ the electric field
between the electrodes is $60\,\textnormal{kV}\cdot\textnormal{m}^{-1}.$
This electric field is not large enough to ionize a neon atom. However,
an electric field of $60\,\textnormal{kV}\cdot\textnormal{m}^{-1}$
at relatively warm temperature is sufficient to emit electrons from
the surface of the cathode lead. Because neon is highly electronegative,
it attracts any free electrons nearby and becomes negatively charged.\cite{electronegativity}
Such case, in which a negatively charged neon atom oscillates in vicinity
of the cathode, is qualitatively explained by the theory presented
in this paper.

\section{Concluding Remarks }

The self-sustained electrode current oscillations in the positive
glow corona can be qualitatively explained by the oscillatory solutions
which is quite naturally obtained from the associated electromagnetic
boundary value problem. To demonstrate this, a simple, DC voltage
biased, plane-parallel plate system with a charged-particle inside
has been considered. The resulting oscillatory solutions for the charged-particle
motion qualitatively explains the observed experimental results. The
remarkable similarities in the waveforms of the self-sustained electrode
current oscillations between the various experiments\cite{corona-discharge-1,corona-discharge-2,corona-discharge-4}
and the prediction from this work indicate that the basic underlying
mechanism behind the self-sustained oscillations in the positive glow
corona involves the kind of push-pull mechanism discussed in Fig.
\ref{fig:3}. \\

\section{Acknowledgments}

The author acknowledges the support for this work provided by Samsung
Electronics Co., Ltd.


\begin{thebibliography}{References}
\bibitem{J. D. Jackson} J. Jackson, Classical Electrodynamics - Third
Edition, Ch. 2 (John Wiley \& Sons, Inc., 1998). 

\bibitem{corona-discharge-1} M. Goldman, A. Goldman, and R. Sigmond,
 Pure \& Appl. Chem. \textbf{57}(9), 1353-1362 (1985).

\bibitem{corona-discharge-2} R. Sigmond,  J. Phys. IV France \textbf{7},
C4-383 (1997)

\bibitem{corona-discharge-3} R. Morrow,  J. Phys. D: Appl. Phys.
\textbf{30}, 3099 (1997). 

\bibitem{corona-discharge-4} Yu. Akishev, M. Grushin, A. Deryugin,
A. Napartovich, M. Pan'kin, and N. Trushkin,  J. Phys. D: Appl. Phys.
\textbf{32}, 2399 (1999). 

\bibitem{corona-discharge-5} N. Allen, M. Abdel-Salam, M. Boutlendj,
I. Cotton, and B. Tan, IET Sci. Meas. Technol.\textit{ }\textbf{1}(2),
103 (2007). 

\bibitem{Cho} S. Cho,  Phys. Plasmas \textbf{19}(3), 033506 (2012).

\bibitem{Japan-al2o3} K. Tamura, Y. Kimura, H. Suzuki, O. Kido, T.
Sato, T. Tanigaki, M. Kurumada, Y. Saito, and C. Kaito, Jpn. J. Appl.
Phys. \textbf{42}, 7489 (2003).

\bibitem{al203} R. Sohal, G. Lupina, O. Seifarth, P. Zaumseil, and
C. Walczyk, Surface Science \textbf{604}, 276 (2010).

\bibitem{IEEE-vacuum} N. Zouache and A. Lefort,  IEEE Trans. Dielectr.
Electr. Insul. \textbf{4}(4), 358 (1997). 

\bibitem{XUV} T. Higashiguchi, H. Terauchi, N. Yugami, T. Yatagai,
W. Sasaki, R. D'Arcy, P. Dunne, and G. O'Sullivan, Appl. Phys. Lett.
\textbf{96}(13), 131505 (2010). 

\bibitem{Bogaerts-DC-glow-discharge} A. Bogaerts, E. Neyts, R. Gijbels,
and Joost. van der Mullen, Spectrochimica Acta Part B\textit{ }\textbf{57}(4),
609 (2002). 

\bibitem{Gyergyek} T. Gyergyek, M. \v{C}er\v{c}ek, M. Stanojevi\'{c},
and N. Jeli\'{c}, J. Phys. D: Appl. Phys. \textbf{27}, 2080 (1994). 

\bibitem{Kuschel-axial-light} T. Kuschel, B. Niermann, I. Stefanovi\'{c},
M. Böke, N. Škoro, D. Mari\'{c}, Z. Petrovi\'{c}, and J. Winter, Plasma
Sources Sci. Technol. \textbf{20}, 065001 (2011). 

\bibitem{field-emission-neon} D. Brandon, Br. J. Appl. Phys. \textbf{14},
474 (1963).

\bibitem{electronegativity} N. Islam and D. Ghosh, J. of Quantum
Information Science \textbf{1}, 135 (2011). \end{thebibliography}
\end{document}